\documentclass[%
 reprint,
 amsmath,amssymb,
 aps,
]{revtex4-2}

\usepackage{graphicx}
\usepackage{dcolumn}
\usepackage{bm}

\begin{document}

\preprint{APS/123-QED}

\title{Vectorial probability loophole in Bell test}

\author{Sheng Feng}
\email{fengsf2a@hust.edu.cn}
\author{Chaoran Tu}

\affiliation{%
School of Electrical and Electronic Information Engineering,\\
 Hubei Polytechnic University, Huangshi, Hubei 435003, China
}%
\affiliation{%
 Center for Fundamental Physics,\\
 Hubei Polytechnic University, Huangshi, Hubei 435003, China
}%

\date{\today}

\begin{abstract}
Exhaustively identifying all loopholes in the Bell test is demanding for interpreting the results of the relevant experiments, since any loophole if not closed can be catastrophic to our understanding of the nonlocal structure of quantum mechanics. Despite a series of recent Bell experiments that claim to be free of loopholes, a united framework with a sound base is still missing to fully recognize all potential loopholes in these experiments, as verified by a recent experiment that has pinned down a detection loophole of a new type in Bell analysis. Here, we reveal another loophole previously unknown in the Bell test through a local theory developed here on the basis of a new mathematical concept of high-dimensional vectorial probability, quantified as a vector with interesting but hidden geometry in the probability space. We show that the statistical property of the stochastic events generated for the Bell test can be well described by the local theory, and, in particular, the correlations of these events that violate Bell's theorem can be closely connected to the geometry of the vectorial probability. To close the loophole, theoretical investigations are highly recommended to search for the statistical nature of the stochastic events in quantum measurements that can distinguish the predictions of quantum mechanics and those of the local theory.
\end{abstract}

\maketitle

\section{Introduction}
Quantum correlations that violate Bell's theorem are invaluable resources for fundamental research \cite{Bell1964,Clauser1969,Clauser1974,horodecki2009,Hirsch2013,Brunner2014,Popescu2014,Giustina2015,Shalm2015,Hensen2015,Deng2018,uola2020,Tendick2020,Atlas2024,Wang2025,Drmota2025,Pettini2025,wang20251} and practical applications in high-tech development \cite{nielsen2000,Barrett2005,Gerhardt2011,Gerhardt2011p,Christensen2013,benenti2019,Zia2023,Liu2024,Cao2024,YLi2025}. Pioneering works \cite{Bell1964,Clauser1969,Freedman1972,Clauser1974,Fry1976,Aspect1982,weihs1998}, which were dedicated to the study of the nonlocal structure of quantum mechanics by exploring these quantum correlations and eventually led to the blossoming of the field of quantum information science \cite{nielsen2000,scarani2009,ladd2010,benenti2019,Arute2019,ma2023,Wang2023,Zapatero2023}, have been celebrated with the 2022 Nobel Prize in Physics. Nonlocality, which claims evidence from the violation of Bell's theorem (Bell violation) \cite{Bell1964,Clauser1969,Clauser1974,Eberhard1993,Giustina2013,Bierhorst2015} observed in Bell experiments \cite{Freedman1972,Fry1976,Aspect1982,weihs1998,Giustina2015,Shalm2015,Hensen2015}, is believed to occupy a central position in quantum mechanics and relevant advances in quantum technology \cite{Brunner2014}. However, the presence of even a loophole in the Bell test endorses a possible explanation of the relevant results by invoking a local theory. As such, an irrefutable test of the nonlocality in the experiments demands a complete account of all possible loopholes, which, however, is still a demanding task to be accomplished.

Currently, to draw a loophole-free conclusion from the relevant results obtained in experiments, physicists commonly consider four main loopholes in the Bell test: the locality loophole \cite{weihs1998}, the detection loophole \cite{Eberhard1993,Giustina2013,Storz2023}, the freedom-of-choice loophole \cite{Hensen2015,Giustina2015,Shalm2015,bigbell2018}, and the post-selection loophole \cite{strekalov1996,santagiustina2024}. Although it has been widely accepted that the theoretical basis for the Bell test is as solid as possible as long as these four loopholes are fully closed in experiments, a recent advance in experiment \cite{tu2026} has shown the existence of a new mechanism other than unfair sampling that can open the detection loophole in the Bell test based on the Clauser-Horne-Shimony-Holt (CHSH) form of the Bell inequality \cite{Clauser1969}. From this result, it follows that the existing theoretical consideration in the Bell test has missed the new mechanism behind the detection loophole. Although most scientists believed that the detection loophole should be completely closed if only the quantum efficiency of the detectors in the experiments is greater than a threshold value $\eta_{\rm th}=82.8\%$, the true value of $\eta_{\rm th}$ is greater than $82.8\%$ and varies as a function of the degree (denoted $S_c$) to which the CHSH inequality is violated due to the detection loophole beyond unfair sampling, according to the new discovery \cite{tu2026}. Then a natural question of fundamental interest is whether there exist other mechanisms that can open unknown loopholes in the Bell test, which is the primary incentive of the current work.
 
Inspired by the experiment that revealed the detection loophole beyond unfair sampling in the Bell test with the CHSH inequality, we first conduct an experiment to uncover another mechanism that can lead to the violation of another form of the Bell inequality, that is, the the Clauser-Horne-Eberhard (CH-E) inequality \cite{Eberhard1993,Bierhorst2015}. This experiment gives an important hint for the existence of a more generic mechanism that can open a loophole in the Bell test with inequality \cite{Svetlichny1987,Hamel2014} or without inequalities \cite{Greenberger1990,Bouwmeester1999,pan2000}. To describe this mechanism in theory, we need to introduce a new concept in mathematics, vectorial probability, as a generalization of the scalar probability from a one-dimensional probability space to higher dimensions. With the concept of vectorial probability, we develop a local theory and show that all known forms of the Bell inequality \cite{Clauser1969,Clauser1974,Eberhard1993,Bierhorst2015} can be violated in the same manner as predicted by quantum mechanics, due to the geometry of vectorial probabilities in a high-dimensional probability space. It is the geometry of the vectorial probabilities that constitutes the mechanism of Bell violation that can open a loophole (namely, the vectorial probability loophole) in the Bell test.

In the next section, we present in detail how to conduct the experiment on the violation of the CH-E inequality by classical correlations of stochastic events. For these events, the experiment can reach an equivalent quantum efficiency equal to unity, a perfect value for the quantum efficiency that is believed to be sufficient to completely close the detection loophole in the Bell test. The mechanism of Bell violation in this experiment can be described by a mathematical model, which suggests that it would be of great interest to generalize the scalar concept of probability to high dimensions to reveal the mechanism that can open a loophole of a new kind in the Bell test with an inequality or without inequalities. In the third section, we provide the conceptual definition of vectorial probability and quantify it as a mathematical vector on which the mathematical operations of relevance can be defined. With the new mathematical concept, a local theory is developed, showing that the length of the vectorial probability locally determines the traditional probability of a stochastic event that occurs per observation (or measurement) or per period of observation (or measurement) time $\tau$. In addition, the joint probability of two or more stochastic events is determined by a product operation (to be defined in the following) of the corresponding vectorial probabilities, each of which governs one of these events.

In section four, we show how the statistical nature of the stochastic events in a Bell test using entangled particles can be described by the local theory with the concept of vectorial probability. In particular, we show how to calculate the joint probability of stochastic events resulting from spin (or polarization) measurement on the system of $m$ entangled particles ($m\ge 2$). These results uncover the existence of a loophole of a new kind in the relevant Bell test, that is, the violation of the Bell inequality can be well explained by the previously unknown geometry of vectorial probabilities in a high-dimensional probability space. In this explanation, the stochastic events in particle detection for spin (polarization) measurement are locally governed by the vectorial probabilities. The fifth section is devoted to discussions on the impact of the current work on the study of quantum physics and relevant fields. To rule out the vectorial probability loophole in the Bell test, further theoretical investigations are highly desirable to discriminate the developed local theory from quantum mechanics regarding their predictions about the violation of Bell's theorem.

\section{Violation of the CH-E inequality by classical correlations}

In Bell's theorem, a series of mathematical inequalities has been derived as a criterion to justify whether the structure of quantum mechanics is nonlocal \cite{Bell1964,Clauser1969,Clauser1974,Eberhard1993,Bierhorst2015}. Among these inequalities, the CHSH inequality \cite{Clauser1969} is of particular interest in the Bell test because it is considered a close variant of the original form of the Bell inequality \cite{Bell1964} that can be used to make quantum mechanical predictions in practical tests. For the CHSH inequality, most scientists believed that the detection loophole can certainly be closed if the quantum efficiency $\eta_{\rm qe}$ of particle detection in Bell experiments is greater than $82.8\%$ \cite{Eberhard1993,Giustina2013}. However, recent progress achieved in an experiment \cite{tu2026} has shown that this loophole can be open even if $\eta_{\rm qe}>82.8\%$, due to a previously unknown mechanism other than unfair sampling in the Bell test.

In what follows, we present an experiment performed by two persons (Alice and Bob) to show the presence of another new loophole in the Bell test based on the CH-E inequality; this loophole is present even if the quantum efficiency of particle detection is perfect. The experiment was conducted using classical devices that created stochastic events with binary results that were determined according to 
\begin{equation} \label{eq:outcfunc}
O_{\pm}(\gamma,\tau)=\mbox{sign}\big(\pm\cos(\gamma+\tau)\big)F\ ,
\end{equation}
where $\gamma$ and $\tau$ are two real numbers, and $F=1$. Specifically, $O_{\pm}(\gamma,\tau)=1$ if $\pm\cos(\gamma+\tau)>0$, and otherwise $O_{\pm}(\gamma,\tau)=-1$ if $\pm\cos(\gamma+\tau)<0$. The probability that the events were produced was unity. However, for a specific result of $O_{+}(\gamma,\tau)$, the probability of event generation was 
\begin{equation} \label{eq:probfuncc}
    P_{\kappa}(\gamma)=\frac{1}{2\pi}\int_0^{2\pi} {\rm d}\tau\ P_{\kappa}(\gamma,\tau),
\end{equation}
in which 
\begin{equation} \label{eq:pprobfunc}
    P_{\kappa}(\gamma,\tau)= |\pm\cos(\gamma+\tau)|^{1/\kappa},
\end{equation}
and $\kappa>0$ is an experimental parameter with a real value that can critically affect the performance of the device in the production of stochastic events. For the result $O_{-}(\gamma,\tau)$, the corresponding probability was $Q_{\kappa}(\gamma)=1-P_{\kappa}(\gamma)$, which ensures the unity probability of event production for both results. Alternatively, the probability of event production can be $P_{\kappa}(\gamma)$ for the result $O_{-}(\gamma,\tau)$, in which case the probability for the result $O_{+}(\gamma,\tau)$ should be $Q_{\kappa}(\gamma)$. In any case, $P_{\kappa}(\gamma)+Q_{\kappa}(\gamma)=1$ and therefore the binary-result events were created with certainty in the experiment. To implement the experiment, however, we need to discretize the expression of the probability given in Eq. (\ref{eq:probfuncc}) for event generation in the following form
\begin{equation} \label{eq:dprobfunc}
    P_{\kappa}(\gamma)=L^{-1}\sum_{j=1}^LP_{\kappa}(\gamma,j\iota),
\end{equation}
where $L>>1$ is an integer and $\iota\equiv2\pi L^{-1}$. 

In the following, we provide an algorithm for the generation of stochastic events in the experiment. With a given real number $\iota$ and a random integer $j$, an event $a$ with binary outcomes is produced at Alice's place with a unity probability; the probability that the event is generated with the result of $O_{-}(\alpha,j\iota)$ is $P_{\kappa}(\alpha,j\iota)$, and the probability for the result $O_{+}(\alpha,j\iota)$ is $Q_{\kappa}(\alpha,j\iota)$. Similarly, another event $b$ with binary outcomes is created at Bob's place with a unity probability as well; the probability of the event generation for the result $O_{+}(\beta,j\iota)$ is $P_{\kappa}(\beta,j\iota)$, and the probability for the result $O_{-}(\beta,j\iota)$ $Q_{\kappa}(\beta,j\iota)$. 

The experiment was carried out with devices and an operation procedure given in \cite{tu2026}. For different values $\kappa$, we conducted the experiment to obtain data for normalized correlations, denoted $C_{AB}(\alpha,\beta;j)$ ($A=\pm,B=\pm$), and single counts, denoted $S_{0B}(\alpha,\beta;j)$ or $S_{A0}(\alpha,\beta;j)$. Here, the subscript $AB$ in $C_{AB}(\alpha,\beta;j)$ means that, in a coincidence count obtained with a random number $j$, Alice created an event $a$ with a result corresponding to $A$ for a parameter value $\alpha$ and, at the same time, Bob created an event $b$ with a result corresponding to $B$ for a parameter value $\beta$. The subscript $0B$ in $S_{0B}(\alpha,\beta;j)$ means that the single count came from the event $b$ created by Bob with a result corresponding to $B$ for a random number $j$, in which case Alice did not produce an event. By analogy, the subscript $A0$ in $S_{A0}(\alpha,\beta;j)$ has a similar meaning. 

With the data acquired in the experiment, violation of the CH-E inequality can be proved as explained in the following. The CH-E form of Bell's inequality reads \cite{Eberhard1993,Bierhorst2015}
\begin{eqnarray}\label{eq:chE}
 S_{\rm e}&=&p_{+-}(\alpha,\beta)-p_{+0}(\alpha,\beta')-p_{0-}(\alpha',\beta)\nonumber\\
 &&-p_{+-}(\alpha',\beta')\le 0\ ,\ \ \ 
\end{eqnarray}
where $p_{+-}(\cdot)$ stands for the joint probability that both events are recorded, with event $a$ having an outcome $+1$ and event $b$ having $-1$; $p_{+0}(\cdot)$ is the probability that the event $a$ is recorded with an outcome $+1$ but without a registered record for $b$, and $p_{0-}(\cdot)$ is the probability that the event $b$ is recorded with an outcome $-1$ but without a record for $a$. 

\begin{figure}[htbp]
\centering
\includegraphics[width=9cm]{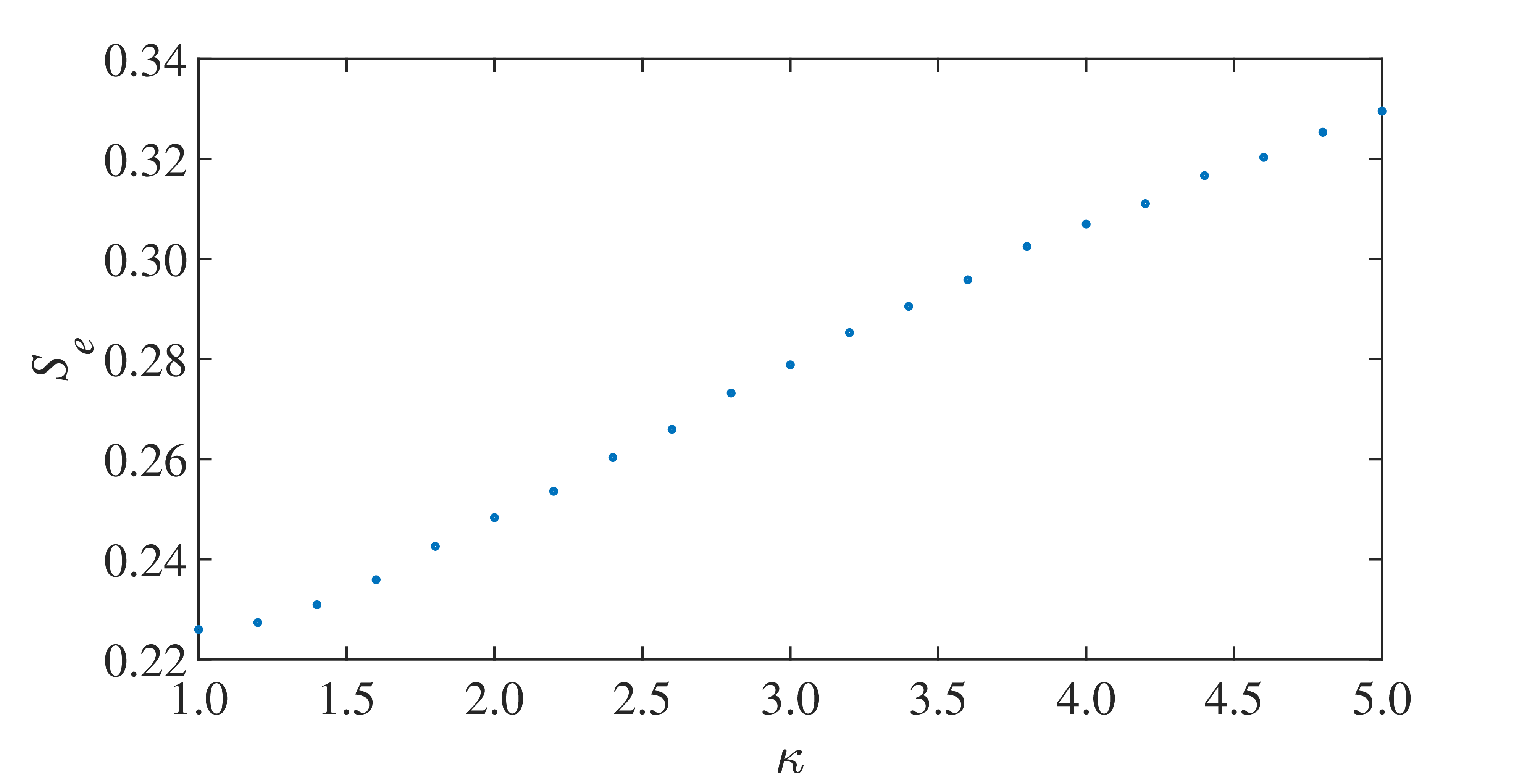}
\caption{Experimental Bell quantity $S_{\rm e}$ versus the parameter $\kappa$. For each data point shown, the number of experimental trials was $L=10^6$. The results show that the CH-E form of Bell's inequality was violated in the experiment with the chosen settings of $\alpha=0$, $\alpha'=\pi/2$, $\beta=0$ and $\beta'=-\pi/2$.}
\label{fig:seeffkexp}
\end{figure}

The value of the term $p_{+-}(\alpha,\beta)$ in inequality (\ref{eq:chE}) was obtained from the data using
\begin{eqnarray}\label{eq:ppmnorm}
&&p_{+-}(\alpha,\beta)=L^{-1}\sum_{j}C_{+-}(\alpha,\beta;j)\ ,
\end{eqnarray}
and the value of the term $p_{+0}(\alpha,\beta')$ in the inequality was obtained by
\begin{eqnarray}\label{eq:ppm0norm}
&&p_{+0}(\alpha,\beta')=L^{-1}\sum_{j}S_{+0}(\alpha,\beta';j)\ .
\end{eqnarray}
The values of other terms in the inequality (\ref{eq:chE}) were calculated in a way similar to Eq. (\ref{eq:ppmnorm}) or Eq. (\ref{eq:ppm0norm}). To evaluate the performance of the event production devices in the experiment, we chose the values of $\alpha$, $\alpha'$, $\beta$, and $\beta'$ with a strategy that maximizes $p_{+-}(\alpha,\beta)$ and minimizes $p_{+-}(\alpha',\beta')$. This strategy was based on the fact that the terms of $p_{+0}(\alpha,\beta')$ and $p_{0-}(\alpha',\beta)$ were zero in the event generation algorithm due to the perfect equivalent quantum efficiency of $100\%$. In this consideration, the parameter settings chosen were $\alpha=\beta=0$ and $\alpha'=-\beta'=\pi/2$, leading to the results presented in Fig. \ref{fig:seeffkexp} that violated the CH-E inequality (that is, $S_{\rm e}>0$).

The above experiment achieved results showing that the CH-E inequality can be violated by classical correlations even if the equivalent quantum efficiency for the generation of stochastic events is $100\%$, which is in sharp contrast to the widely accepted belief that this inequality can be violated only if the efficiency is lower than $66.7\%$ \cite{Eberhard1993}. As such, a reasonable explanation for the reported results is highly desired. In the following, we show how the violation of the CH-E inequality in the above experiment can be explained on the basis of the mathematical concept of vectorial probability. To this end, we define two vector quantities, respectively, as
\begin{eqnarray} \label{eq:vecdef0}
    \mathbf{p}_\alpha&=&\big[P_\alpha O_-,Q_\alpha O_+,P_\alpha O_-,Q_\alpha O_+\big] \nonumber\\
    \mathbf{p}_\beta&=&\big[P_\beta O_+,Q_\beta O_-,Q_\beta O_-,P_\beta O_+\big]\ ,
\end{eqnarray}
where $P_\alpha\equiv P_\kappa(\alpha,\tau)$, $P_\beta\equiv P_\kappa(\beta,\tau)$, and $O_\pm\equiv O_\pm(\gamma,\tau)$ with $\gamma=\alpha\ {\rm or}\ \beta$. In our explanation, the production of the stochastic events by Alice or Bob was ruled by the two vectorial probabilities defined in Eqs. (\ref{eq:vecdef0}). As such, the joint probability of producing stochastic events by Alice and Bob in the experiment can be calculated as
\begin{eqnarray} \label{eq:corrck}
C_{\kappa}(\alpha,\beta)&=&\mathbf{p}_\alpha \cdot\mathbf{p}_\beta\nonumber\\
&=&\sum_{j=1}^L\Big[P_{\kappa}(\alpha,j\iota)\ O_{-}(\alpha,j\iota)\nonumber\\
&&+\big(Q_{\kappa}(\alpha,j\iota)\big)O_{+}(\alpha,j\iota)\Big]_a\nonumber\\
&&\times\Big[P_{\kappa}(\beta,j\iota)\ \ O_{+}(\beta,j\iota)\nonumber\\
&&+\big(Q_{\kappa}(\beta,j\iota)\big)\ O_{-}(\beta,j\iota)\Big]_b.\ \  \ 
\end{eqnarray}
It is not difficult to see that the above expression (\ref{eq:corrck}) exactly describes the algorithm for the joint production of the stochastic events in the experiment. We suspect that the dot-product of the two vectors, $\mathbf{p}_\alpha$ and $\mathbf{p}_\beta$, is associated with some previously unknown geometry owned by the vectorial probabilities in a probability space, and the violation of the CH-E inequality can be attributed to this hidden geometry that governed the correlation of the stochastic events created by Alice and Bob. Inspired by this insight, in what follows, we provide a more generic definition of the concept of vectorial probability so that a local theory can be developed to explain violation of the Bell's inequality in all known forms and even violation of Bell's theorem without any inequalities. 

\section{Theory of vectorial probability}

The concept of vectorial probability is a natural generalization of the traditional scalar probability from a one-dimensional space (which here is called the probability space) to higher dimensions. To understand why and how the new concept is introduced in theory, consider first the game of coin flip, in which a coin is thrown into the air, causing it to rotate quickly edge-over-edge. In the end, the coin will land with the heads or tails facing up and the outcomes of the flips appear with some (scalar) probabilities, respectively, denoted $p_h$ and $p_t$ to quantify the average number of each outcome per toss. Typically, $p_{h,t}\ge 0$, $p_h+p_t=1$, and particularly $p_h=p_t=0.5$ should be expected for the fairness of the game. One must stress that in the mentioned scenario, the word ``probability" is referred to as the average number of an outcome per toss. In this study, we examine the game from a more generic perspective and suppose that a coin toss is performed randomly many times during a certain period of time $T$, for which the term ``probability" (denoted $p$, a scalar) is used to describe the likelihood of an event with a unary result (either heads or tails, but not both) per observation (measurement) time $\tau$ or per observation (measurement). In the former case, the value of $\tau$ can be appropriately chosen such that $0\le p\le 1$.

The probability $p$, introduced above as a scalar quantity in a one-dimensional probability space, can be generalized to higher dimensions, as follows. Suppose that two stochastic events ($a_1$ and $a_2$) occur with unary results at two places at a distance. Let us use $p_{a_1}\ge 0$ and $p_{a_2}\ge 0$ respectively to denote the probabilities of the two events during the time $\tau$. If the events are not correlated, the joint probability $p_{a_1a_2}$ that both events take place during the time $\tau$ should be $p_{a_1a_2}=p_{a_1} p_{a_2}$. However, if the two events occur with correlations, the concept of vectorial probability may be needed to quantitatively describe the joint probability of the events, such as particle detection events in the Bell test \cite{Giustina2015,Shalm2015,Hensen2015}.

Vectorial probability, denoted by $\mathbf{p}(\xi)$, is defined as a vector in an $N$-dimensional probability space $S$ ($N$ is an integer, and $N\ge2$). The parameter $\xi$ represents a single (discrete or continuous) variable or a set that can locally dictate the vector. For a stochastic event ruled by a vectorial probability $\mathbf{p}(\xi)$, the length of the vector, $\lVert\mathbf{p}(\xi)\rVert$, in the space $S$ is mapped to the (scalar) probability $0\le p(\xi)\le 1$ in a one-dimensional space for the same event, that is, $p(\xi)=\lVert\mathbf{p}(\xi)\rVert$. For two stochastic events $a_1$ and $a_2$, their joint probability $p_{a_1a_2}$ is calculated as the dot-product of the vectorial probabilities of the corresponding events $\mathbf{p}_{a_1}(\xi_{1})$ and $\mathbf{p}_{a_2}(\xi_{2})$,
\begin{eqnarray} \label{eq:defp}
    p_{a_1a_2}(\xi_{1},\xi_{2}) &=& \mathbf{p}_{a_1}(\xi_{1}) \cdot \mathbf{p}_{a_2}(\xi_{2}) \nonumber\\
    &=&\lVert\mathbf{p}_{a_1}(\xi_{1})\rVert\  \lVert\mathbf{p}_{a_2}(\xi_{2})\rVert \cos \Omega(\xi_{1},\xi_{2})\nonumber\\
    &=&p_{a_1}(\xi_{1})p_{a_2}(\xi_{2}) \cos \Omega(\xi_{1},\xi_{2})\ ,
\end{eqnarray}
where $\Omega(\xi_{1},\xi_{2})$ is the angle between the vectors $\mathbf{p}_{a_1}(\xi_{1})$ and $\mathbf{p}_{a_2}(\xi_{2})$ in the space $S$, and $p_{a_1}(\cdot)$ and $p_{a_2}(\cdot)$ are the corresponding probabilities (scalar) of the two events. From $p_{a_1a_2}(\xi_{1},\xi_{2}) \ge 0$ it follows that $\cos \Omega(\xi_{1},\xi_{2}) \ge 0$, which involves an acute angle or a right angle between $\mathbf{p}_{a_1}(\xi_{1})$ and $\mathbf{p}_{a_2}(\xi_{2})$.

Suppose that the probability space $S$ is spanned by a set of orthogonal unity base vectors $\hat{s}_j$ ($j=1, 2, ..., N$, and $N\ge 2$), then the vectorial probability $\mathbf{p}(\xi)$ may be written as a summation of $N$ components,
\begin{equation} \label{eq:decom}
    \mathbf{p}(\xi) = \sum_{j=1}^N p_j(\xi) \ \hat{s}_j\ ,
\end{equation}
in which $p_{j}(\xi)$ (a real number for the moment, but can be a complex number in general, as will be shown later) denotes the projected component of $\mathbf{p}(\xi)$ along the $\hat{s}_{j}$ direction, i.e., $p_j(\xi)=\mathbf{p}(\xi) \cdot \hat{s}_j$ ($j=1,...,N$). For brevity, the vectorial probability $\mathbf{p}(\xi)$ of Eq. (\ref{eq:decom}) can be re-expressed as $\mathbf{p}(\xi)\equiv\left[p_1(\xi), p_2(\xi), ..., p_N(\xi)\right]$. Note that $p_j(\xi)$ can have a negative value and therefore may not necessarily be interpreted to have the meaning of traditional probability, although Eq. (\ref{eq:decom}) can be used as a tool to calculate the joint probability of the two events $a_1$ and $a_2$. One may plug Eq. (\ref{eq:decom}) into Eq. (\ref{eq:defp}), arriving at 
\begin{equation} \label{eq:jointp}
    p_{a_1a_2}(\xi_{1},\xi_{2}) = \sum_{j=1}^N p_{a_1,j}(\xi_{1})\ p_{a_2,j}(\xi_{2})\ ,
\end{equation}
wherein $p_{a_1,j}(\xi_{1})=\mathbf{p}_{a_1}(\xi_{1}) \cdot \hat{s}_j$ and $p_{a_2,j}(\xi_{2})=\mathbf{p}_{a_2}(\xi_{2}) \cdot \hat{s}_j$ are respectively the projected components of $\mathbf{p}_{a_1}(\xi_1)$ and $\mathbf{p}_{a_2}(\xi_2)$ along the $\hat{s}_{j}$ direction. 

For stochastic events, $a_1$, ..., and $a_m$ ($m\ge 3$), to calculate the joint probability it is necessary to introduce a new mathematical operation for the corresponding vectorial probabilities $\mathbf{p}_{a_k}(\xi_k)$ ($k=1,2,...,m$). To this aim, define a mathematical operation for vectors, namely the successive-dot product (s-dot product for brevity),
\begin{eqnarray} \label{eq:udp}
&&\mathbf{p}_{a_1}(\xi_1)\cdot ... \cdot \mathbf{p}_{a_k}(\xi_k)\cdot ... \cdot \mathbf{p}_{a_m}(\xi_m)\nonumber\\
&=&\sum_{j=1}^N p_{a_1,j}(\xi_1)... p_{a_k,j}(\xi_k)... p_{a_m,j}(\xi_m)\ ,
\end{eqnarray} 
which made use of Eq. (\ref{eq:decom}) to expand $\mathbf{p}_{a_k}(\xi_k)$ ($k=1,2,...,m$) as a summation of $N$ components. Again, $p_{a_k,j}(\xi_k)$ may not necessarily be interpreted as the probability of an event but is useful to calculate the joint probability of the involved stochastic events, 
\begin{eqnarray} \label{eq:jointabc}
 &&p_{a_1...a_k...a_m}(\xi_1,...,\xi_k,...,\xi_m)\nonumber\\
&=&\sum_{j=1}^N p_{a_1,j}(\xi_1)... p_{a_k,j}(\xi_k)... p_{a_m,j}(\xi_m)\ .
\end{eqnarray}

\subsection{Coordinated stochastic events}

Now consider a special situation where the coin flip game is played by a person who can watch the play through a mirror and records both the real outcomes (heads or tails) and the ones observed through the mirror. Mathematically, there exist two stochastic events simultaneously occurring in this scenario, a real event and a virtual one in the mirror. Let $p_r$ and $p_v$, respectively, denote the probabilities of the two events, then $p_r=p_v$ and their joint probability is $p_{rv}=p_r=p_v$. Please note that the joint probability cannot be calculated as $p_{rv}=p_rp_v$ in this case because the two events are highly coordinated to produce outcomes that are exact duplicates of each other during any observation (observation) time.

In a probability space $S$ with a higher dimension, two stochastic events, $a_1$ and $a_2$, are said to be coordinated if and only if (1) $\mathbf{p}_{a_1}(\xi_{1})=\mathbf{p}_{a_2}(\xi_{2})$ when ${\xi_{2}=\xi_{1}}$, that is, $ \mathbf{p}_{a_1}(\xi_{1})=\mathbf{p}_{a_2}(\xi_{2})|_{\xi_{2}=\xi_{1}}$; and (2) their joint probability for $\xi_{2}=\xi_{1}$ reads 
\begin{eqnarray} \label{eq:defcp}
&&p_{a_1a_2}(\xi_{1},\xi_{2})|_{\xi_{2}=\xi_{1}}=p_{a_1}(\xi_1)=p_{a_2}(\xi_2)|_{\xi_{2}=\xi_{1}}\ .
\end{eqnarray}
From Eq. (\ref{eq:defcp}) it follows that 
\begin{eqnarray} \label{eq:defcpn}
&&p_{a_1a_2}(\xi_{1},\xi_{2})|_{\xi_{2}=\xi_{1}}\nonumber\\
&=&p^{-1}_{a_1}(\xi_1)\ p_{a_1}(\xi_{1})\ p_{a_2}(\xi_{1}) \nonumber\\
&=&p^{-1}_{a_1}(\xi_1)\ \lVert\mathbf{p}_{a_1}(\xi_{1})\rVert\  \lVert\mathbf{p}_{a_2}(\xi_{1})\rVert\nonumber\\
&=&p^{-1}_{a_1}(\xi_1)\ \mathbf{p}_{a_1}(\xi_{1}) \cdot \mathbf{p}_{a_2}(\xi_{2})|_{\xi_{2}=\xi_{1}} \ ,
\end{eqnarray}
in which the calculation in the last step made use of $\Omega(\xi_{1},\xi_{2})|_{\xi_{2}=\xi_{1}}=0$ because $ \mathbf{p}_{a_1}(\xi_{1})=\mathbf{p}_{a_2}(\xi_{2})|_{\xi_{2}=\xi_{1}}$. 

To generalize Eq. (\ref{eq:defcpn}) to cover the more generic case in which $\xi_{2}\ne\xi_{1}$, one can decompose the vectorial probability $\mathbf{p}_{a_2}(\xi_{2})$ of the event $a_2$ into two parts, one part parallel to $\mathbf{p}_{a_2}(\xi_{1})$ and the other orthogonal to it, as follows,
\begin{eqnarray} \label{eq:decomvp}
\mathbf{p}_{a_2}(\xi_{2})&=&\zeta\ \mathbf{p}_{a_2}(\xi_{1}) + \Big[\mathbf{p}_{a_2}(\xi_{2})- \zeta\  \mathbf{p}_{a_2}(\xi_{1})\Big]\nonumber\\
&=&\zeta\ \mathbf{p}_{a_1}(\xi_{1}) + \Big[\mathbf{p}_{a_2}(\xi_{2})- \zeta\  \mathbf{p}_{a_1}(\xi_{1})\Big]
\ ,\ 
\end{eqnarray}
wherein $\zeta={p}^{-1}_{a_1}(\xi_{1})\ {p}_{a_2}(\xi_{2})\ \cos\Omega(\xi_{1},\xi_{2})$. It is not difficult to prove 
\begin{equation} \label{eq:porth}
\mathbf{p}_{a_1}(\xi_{1})\cdot\Big[\mathbf{p}_{a_2}(\xi_{2})- \zeta\  \mathbf{p}_{a_1}(\xi_{1})\Big]=0\ .    
\end{equation}
Therefore, the first term of Eq. (\ref{eq:decomvp}) via event coordination corresponds to the production of events with outcomes that are duplicates of part of the outcomes of the event $a_1$, described by $\mathbf{p}_{a_1}(\xi_{1})$, while the second does not contribute to the joint probability. Consequently, the joint probability of the two events for $\xi_{2}\ne\xi_{1}$ is reduced by a factor of $\zeta$ compared to that of Eq. (\ref{eq:defcpn}), that is, 
\begin{eqnarray} \label{eq:jointg1}
p_{a_1a_2}(\xi_{1},\xi_{2})|_{\xi_2\ne\xi_1}
&=&p^{-1}_{a_1}(\xi_{1})\ \mathbf{p}_{a_1}(\xi_{1}) \cdot \big[\zeta\mathbf{p}_{a_2}(\xi_{1})\big] \nonumber\\
&=&p^{-1}_{a_1}(\xi_{1})\ \mathbf{p}_{a_1}(\xi_{1}) \cdot \mathbf{p}_{a_2}(\xi_{2}) \ ,\ \ 
\end{eqnarray}
where the last-step calculation invoked Eqs. (\ref{eq:decomvp}) and (\ref{eq:porth}). Similarly, one may also obtain
\begin{eqnarray} \label{eq:jointg2}
p_{a_1a_2}(\xi_{1},\xi_{2})|_{\xi_2\ne\xi_1}
&=&p^{-1}_{a_2}(\xi_{2})\ \mathbf{p}_{a_1}(\xi_{1}) \cdot \mathbf{p}_{a_2}(\xi_{2}) \ .
\end{eqnarray}
With Eq. (\ref{eq:jointg1}) or (\ref{eq:jointg2}), we have a formula generalized from Eq. (\ref{eq:defcpn}) for calculating the joint probability of two coordinated events, each locally described by their own vectorial probability.

Next, we prove an interesting property of the vectorial probabilities of two coordinated events, that is, their lengths are constant independent of the parameter $\xi_1$ or $\xi_2$, which means that events $a_1$ and $a_2$ occur with constant (scalar) probabilities. To this aim, one can show without difficulty, with Eqs. (\ref{eq:jointg1}) and (\ref{eq:jointg2}), that $p_{a_1}(\xi_{1})=p_{a_2}(\xi_{2})$ for $\xi_{1}\ne \xi_2$ and therefore $p_{a_2}(\xi_{2})=p_{a_2}(\xi_{1})$ since $p_{a_2}(\xi_{1})=p_{a_1}(\xi_{1})$. Following the same reasoning, we can obtain $p_{a_1}(\xi_{2})=p_{a_1}(\xi_{1})$, proving that the length of the vectorial probability $\mathbf{p}_{a_1}(\xi_{1})$ is constant. In addition, we can also obtain $\zeta=\cos\Omega(\xi_{1},\xi_{2})$ and therefore $0\le\zeta\le 1$.

Eq. (\ref{eq:jointg1}) or (\ref{eq:jointg2}) can be generalized to the case of more coordinated events. For stochastic events, $a_1$, ..., and $a_m$ ($m\ge 3$), they are coordinated if and only if (1) $\mathbf{p}_{a_1}(\xi_{1})=...=\mathbf{p}_{a_m}(\xi_{m})|_{\xi_{m}=\xi_{1}}$; and (2) their joint probability for $\xi_{m}=...=\xi_{1}$ reads 
\begin{eqnarray} \label{eq:defcpm}
&&p_{a_1...a_m}(\xi_{1},...,\xi_{m})|_{\xi_{m}=...=\xi_{1}}\nonumber\\
&=&p_{a_1}(\xi_1)=...=p_{a_m}(\xi_m)|_{\xi_{m}=\xi_{1}}.
\end{eqnarray}
For the generic case of $\xi_{k_1}\ne\xi_{k_2}$ ($k_1\ne k_2$ and $m\ge k_1,k_2\ge 1$), $p_{a_1}(\xi_{1})=...=p_{a_m}(\xi_{m})$ is expected and the joint probability of these coordinated events reads
\begin{eqnarray} \label{eq:jointabccu}
 &&p_{a_1...a_k...a_m}(\xi_1,...,\xi_k,...,\xi_m)\nonumber\\
 &=&p^{-m+1}_{a_1}(\xi_{1})\ \mathbf{p}_{a_1}(\xi_1)\cdot ... \cdot \mathbf{p}_{a_k}(\xi_k)\cdot...\cdot \mathbf{p}_{a_m}(\xi_m)\nonumber\\
 &=&... ...\nonumber\\
 &=&p^{-m+1}_{a_m}(\xi_{m})\ \mathbf{p}_{a_1}(\xi_1)\cdot ... \cdot \mathbf{p}_{a_k}(\xi_k)\cdot...\cdot \mathbf{p}_{a_m}(\xi_m)\ .\ \ \ \
\end{eqnarray}
We will show below that a local model can be constructed based on Eq. (\ref{eq:jointabccu}) to reproduce the predictions of quantum theory about Bell's theorem when $m$ entangled particles ($m\ge 3$) are involved.

\subsection{Correlations between stochastic events}
Intuitively, the correlation between stochastic events ($a_1$, ..., $a_k$, ..., and $a_m$) with unary outcomes can be defined to be equal to their joint probability. When these events are, respectively, governed by $\mathbf{p}_{a_1}(\xi_1)$, ..., $\mathbf{p}_{a_k}(\xi_k)$, ..., and $\mathbf{p}_{a_m}(\xi_m)$ in the space $S$, their correlation reads
\begin{eqnarray} \label{eq:corrabcp}
&&C_{a_1...a_k...a_m}(\xi_1,...,\xi_k,...,\xi_m)\nonumber\\
&=& p_{a_1...a_k...a_m}(\xi_1,...,\xi_k,...,\xi_m) \nonumber\\
&=&C_{u}\ \mathbf{p}_{a_1}(\xi_1)\cdot ... \cdot \mathbf{p}_{a_k}(\xi_k)\cdot...\cdot \mathbf{p}_{a_m}(\xi_m)
\ ,\ 
\end{eqnarray}
where $C_{u}=p^{-m+1}_{a_1}(\xi_{1})$ if these events are coordinated and $C_{u}=1$ if not. For stochastic events with binary results, the case is more complex and needs further examination.

Reconsider the coin flip game that is played randomly many times during a certain period of time $T$. This time, we want to describe the likelihood of an event (with binary results of heads and tails) per observation (measurement) time $\tau$ or per observation (measurement). To that end, the binary results are labeled $\pm 1$, e.g. the heads as $+1$ and the tails $-1$; the corresponding (scalar) probabilities are respectively denoted $p_{\pm 1}$ ($0\le p_{\pm 1}\le 1$). Then we can define negative probability $q_{-1}=-p_{-1}$ in the sense that the coin is landed with the result $-1$ (that is, it is flipped upside down) with the (traditional) probability of $|q_{-1}|=p_{-1}$ (here $|\cdot|$ stands for the absolute value of a real number). After introducing the concept of negative probability, the game can be described with a unified quantity of probability $q_A=Ap_A$ ($A=\pm 1$) in a one-dimensional probability space, having a positive or negative value. As such, $q_{+1}>0$ has the traditional meaning of probability $q_{+1}=p_{+1}$ for the result $+1$, and $q_{-1}<0$ means that the result $-1$ appears with the conventional probability of $|q_{-1}|=-q_{-1}=p_{-1}$. 

In an $N$-dimensional space $S$, a vectorial probability that describes a stochastic event with result $\pm1$ is denoted $\mathbf{p}_{\pm1}(\xi)$, and the corresponding (scalar) probability of the same event is $p_{A}=\lVert\mathbf{p}_{A}(\xi)\rVert$ ($A=\pm 1$). Because an event cannot occur simultaneously with the two results of $+1$ and $-1$, it is reasonable to expect $\mathbf{p}_{+1}(\xi)\cdot \mathbf{p}_{-1}(\xi)=0$. Then the joint probability of two uncoordinated events ($a_1$ and $a_2$) with outcomes, respectively $A_1$ and $A_2$ ($A_1=\pm 1$, $A_2=\pm 1$), can be calculated as a dot-product of two vectorial probabilities, 
\begin{eqnarray} \label{eq:defpa12}
   &&p_{a_1a_2,A_1A_2}(\xi_{1},\xi_{2})\nonumber\\
   &=& \mathbf{p}_{a_1,A_1}(\xi_{1}) \cdot \mathbf{p}_{a_2,A_2}(\xi_{2}) \nonumber\\
   &=&\lVert\mathbf{p}_{a_1,A_1}(\xi_{1})\rVert\  \lVert\mathbf{p}_{a_2,A_2}(\xi_{2})\rVert \cos \Omega(\xi_{1},\xi_{2})\nonumber\\
   &=&p_{a_1,A_1}(\xi_{1})\ p_{a_2,A_2}(\xi_{2})\ \cos \Omega(\xi_{1},\xi_{2})\ ,
\end{eqnarray}
where $\Omega(\xi_{1},\xi_{2})$ again is the angle between the vectors $\mathbf{p}_{a_1,A_1}(\xi_{1})$ and $\mathbf{p}_{a_2,A_2}(\xi_{2})$. Plugging Eq. (\ref{eq:decom}) in Eq. (\ref{eq:defpa12}), one may arrive at 
\begin{eqnarray}\label{eq:jointpa12}    
p_{a_1a_2,A_1A_2}(\xi_{1},\xi_{2})
&=& \sum_{j=1}^N p_{a_1,A_1;j}(\xi_{1})\ p_{a_2,A_2;j}(\xi_{2})\ .
\end{eqnarray} 
In the notation of the unified quantity of probability, $q_{+1}=\lVert\mathbf{p}_{+1}(\xi)\rVert=p_{+1}$ for the result $+1$ and $q_{-1}=-\lVert\mathbf{p}_{-1}(\xi)\rVert=-p_{-1}$ for the result $-1$. As such, for the (unified) joint probability, we have
\begin{eqnarray}\label{eq:defq12}
&&q_{a_1a_2,A_1A_2}(\xi_1,\xi_2)\nonumber\\
&=&q_{a_1,A_1}(\xi_{1})\ q_{a_2,A_2}(\xi_{2})\ \cos \Omega(\xi_{1},\xi_{2})\nonumber\\
&=& A_1A_2\   p_{a_1a_2,A_1A_2}(\xi_1,\xi_2),
\end{eqnarray}
in which $q_{a_1,A_1}(\xi_{1})=A_1p_{a_1,A_1}(\xi_{1})$ and $q_{a_2,A_2}(\xi_{2})=A_2p_{a_2,A_2}(\xi_{2})$. The correlation of the two events is defined as
\begin{eqnarray} \label{eq:corr12q}
&&C_{a_1a_1}(\xi_1,\xi_2)\nonumber\\
&=&\sum_{A_1,A_2=-1}^{+1} q_{a_1a_2,A_1A_2}(\xi_1,\xi_2)\nonumber\\
&=& \sum_{A_1,A_2=-1}^{+1}A_1A_2  \Big[\mathbf{p}_{a_1,A_1}(\xi_1)\cdot \mathbf{p}_{a_2,A_2}(\xi_2)\Big]\ .\ \ 
\end{eqnarray}

Now consider correlations between two coordinated stochastic events, $a_1$ and $a_2$, with binary results. The two events are coordinated if and only if (1) $ \mathbf{p}_{a_1,A_1}(\xi_{1})=\mathbf{p}_{a_2,A_2}(\xi_{2})|_{\xi_{2}=\xi_{1}}$; and (2) their joint probability for $\xi_{2}=\xi_{1}$ reads 
\begin{eqnarray} \label{eq:defcpbi}
&&p_{a_1a_2,A_1A_2}(\xi_{1},\xi_{2})|_{\xi_{2}=\xi_{1}}\nonumber\\
&=&p_{a_1,A_1}(\xi_1)\nonumber\\
&=&p_{a_2,A_2}(\xi_2)|_{\xi_{2}=\xi_{1}}\ .
\end{eqnarray}
It is not difficult to show that the joint probability of the two events for $\xi_2\ne \xi_1$ is
\begin{eqnarray} \label{eq:jointqco}
&&p_{a_1a_2,A_1A_2}(\xi_{1},\xi_{2})|_{\xi_2\ne\xi_1}\nonumber\\
&=&p^{-1}_{a_1,A_1}(\xi_{1})\ \mathbf{p}_{a_1,A_1}(\xi_{1}) \cdot \mathbf{p}_{a_2,A_2}(\xi_{2}) \nonumber\\
&=&p^{-1}_{a_2,A_2}(\xi_{2})\ \mathbf{p}_{a_1,A_1}(\xi_{1}) \cdot \mathbf{p}_{a_2,A_2}(\xi_{2})\ ,\ \ 
\end{eqnarray}
where $p_{a_1,A_1}(\xi_{1})=p_{a_2,A_2}(\xi_{2})$ holds for $\xi_1\ne \xi_2$, which is similar to the case of coordinated events with unary outcomes. As such, given Eq. (\ref{eq:defcpbi}), the probability of each event is also independent of the parameters $\xi_{1,2}$, that is, $p_{a_1,A_1}(\xi_1)=p_{a_1,A_1}(\xi_2)$ and $p_{a_2,A_2}(\xi_1)=p_{a_2,A_2}(\xi_2)$ even if $\xi_1\ne \xi_2$. From Eqs. (\ref{eq:defcpbi}) and (\ref{eq:jointqco}), one has the correlation of the two coordinated events,
\begin{eqnarray} \label{eq:corr12qco}
&&C_{a_1a_1}(\xi_1,\xi_2)\nonumber\\
&=&\sum_{A_1,A_2=-1}^{+1} q_{a_1a_2,A_1A_2}(\xi_1,\xi_2)\nonumber\\
&=& C_b\sum_{A_1,A_2=-1}^{+1}A_1A_2  \Big[\mathbf{p}_{a_1,A_1}(\xi_1)\cdot \mathbf{p}_{a_2,A_2}(\xi_2)\Big]\ ,\ \ 
\end{eqnarray}
where $C_b=p^{-1}_{a_1,A_1}(\xi_{1})$.

For more stochastic events, $a_1$, ..., and $a_m$ ($m\ge 3$), with binary outcomes of $A_1$, ..., and $A_m$, a generalized correlation function can be obtained similar to Eqs. (\ref{eq:corr12q}) and (\ref{eq:corr12qco}), 
\begin{eqnarray} \label{eq:corrabcq}
&&C_{a_1...a_k...a_m}(\xi_1,...,\xi_k,...,\xi_m)\nonumber\\
&=& \sum_{A_1,...,A_m=-1}^{+1} q_{a_1...a_k...a_m,A_1...A_k...A_m}(\xi_1,...,\xi_k,...,\xi_m) \nonumber\\
&=&  C_{bm}\sum_{A_1,...,A_m=-1}^{+1}A_1...A_k...A_m \nonumber\\
&\times& \Big[\mathbf{p}_{a_1,A_1}(\xi_1)\cdot ...\cdot \mathbf{p}_{a_k,A_k}(\xi_k)\cdot...\cdot \mathbf{p}_{a_m,A_m}(\xi_m)\Big]\ ,\ \ \ 
\end{eqnarray}
in which $C_{bm}=p^{-m+1}_{a_1,A_1}(\xi_{1})$ when these events are coordinated and $C_{ub}=1$ when they are not.

\section{Application of the theory of Vectorial probability to Bell test}

Quantum mechanics has predicted many singular physical phenomena, e.g., Bell violation \cite{Bell1964,Clauser1969} and quantum entanglement \cite{epr}, in the microscopic world. Bell violation has been verified by quantum correlations of stochastic events in Bell experiments \cite{Freedman1972,Aspect1982,weihs1998,Giustina2015,Shalm2015,Hensen2015} that have been proposed and performed following Bell's seminal work \cite{Bell1964}. If no loophole is open in the Bell test, the results of the Bell violation can be claimed as evidence of the nonlocality of quantum mechanics according to Bell's theorem \cite{Bell1964,Clauser1969,Clauser1974,Eberhard1993,Bierhorst2015}. However, exhaustively identifying in theory all loopholes in the Bell test requires a unified framework with solid ground, and constructing such a framework is still necessary. This point gains support from a recent advance in an experiment that has recognized a previously unknown detection loophole in the Bell test \cite{tu2026}. In what follows, we provide another support for the point by showing the presence of a new loophole (namely, the vectorical probability loophole) in the Bell test. To this aim, we apply the theory of vectorial probability to the Bell test and show that violation of Bell's theorem by the correlations of entangled particles can be well explained by the local theory developed above.

\subsection{Bell violation with two entangled particles predicted by quantum mechanics}

Consider first a Bell experiment conducted with two entangled electrons $e_a$ and $e_b$ in a singlet state \cite{Bell1964,Hensen2015}, a quantum state in which the electron spins are perfectly anticorrelated and the total angular momentum (spin) of the whole system is zero. Suppose each particle is sent into a black box where the electron spin is measured by a device following the instructions by Alice or Bob, who decides the direction along which the spin is measured. Each chosen direction may be characterized by a local parameter $\alpha$ or $\beta$, an azimuthal angle in the spatial coordinate system. 

In the experiment, the accomplishment of a spin measurement of an electron is signaled by the detection of the electron with a certain device; in other words, the spin measurement produces stochastic events of electron detection. Suppose that local spin measurements on electrons $e_a$ and $e_b$ produce stochastic particle detection events (events $a$ and $b$) with the results of spin up (labeled +1) or spin down (-1) along the measurement directions, respectively. Quantum theory adopts the scalar concept of conventional probability, which is defined as the likelihood of events per observation (measurement), and predicts that the joint probability $p'_{ab,AB}$ of the particle detection events occurring in the joint spin measurement with the result of $A$ ($A=\pm 1$) for Alice and $B$ ($B=\pm 1$) for Bob is
\begin{eqnarray} \label{eq:jointq}
p'_{ab,AB}(\alpha,\beta)|_{A=-B}&=&2^{-1}\cos^2(\alpha/2-\beta/2)\ , \ \ \ \mbox{or}\nonumber\\
p'_{ab,AB}(\alpha,\beta)|_{A=B}\ \ &=&2^{-1}\sin^2(\alpha/2-\beta/2)\ .
\end{eqnarray}
Using Eq. (\ref{eq:corr12qco}) or (\ref{eq:corrabcq}), the following correlation can be trivially obtained between the local particle detection events, respectively, in the spin measurements by Alice and Bob in the experiment,
\begin{eqnarray}\label{eq:coscorr}
    C_{ab}(\alpha,\beta)
   &=&\sum_{A,B=-1}^{+1} A\ B \ p'_{ab,AB}(\alpha,\beta)\nonumber \\ 
    &=&-\cos(\alpha-\beta)\ .
\end{eqnarray} 
According to the process provided by Bell \cite{Bell1964}, the correlation of Eq. (\ref{eq:coscorr}) violates Bell's theorem, as shown below.

In history Bell was the first one to try to establish a connection of a local probabilistic model to the correlation between two stochastic events given by Eq. (\ref{eq:coscorr}). Specifically, in this model, a correlation function (expectation function, in Bell's language) is defined as \cite{Bell1964}
\begin{equation} \label{eq:expectation}
    C(\alpha,\beta)=\int \mbox{d}\chi\ \rho(\chi) A(\chi,\alpha)B(\chi,\beta),
\end{equation}
where $A(\chi,\alpha)=\pm 1$ and $B(\chi,\beta)=\pm 1$ are (probabilistically) deterministic functions of $\chi$, $\alpha$ and $\beta$, and $\rho(\chi)\ge 0$ denotes the probability distribution of $\chi$ in the sense that $\int \mbox{d}\chi\ \rho(\chi)=1$. An implicit but vital assumption for the model is that the result $B$ for the event $b$ does not depend on the value of $\alpha$, nor $A$ on $\beta$ \cite{Bell1964}, which is what the term ``a local model" means. For the two particles in the entangled state in which $A(\chi,\alpha)=-B(\chi,\alpha)$, Bell showed that their correlation violates what is referred to as the original form of Bell's inequality, 
\begin{equation}\label{eq:bell}
    S_b=|C(\alpha,\beta)-C(\alpha,\gamma)|- C(\beta,\gamma)-1\le 0\ , 
\end{equation}
in which $\gamma$ has the same meaning as $\alpha$ or $\beta$. 

Following Bell's arguments, if the system composed of two entangled electrons is local in the sense that ``the result $B$ for the electron $e_b$ does not depend on the value of $\alpha$, nor $A$ on $\beta$", the stochastic events produced in spin measurement must have correlation that obeys the Bell inequality (\ref{eq:bell}). However, the correlation (\ref{eq:coscorr}) predicted by quantum theory indeed violates Bell's inequality (\ref{eq:bell}), which can be trivially proved by setting $\alpha=0$, $\beta=\pi/4$, and $\gamma=3\pi/4$, leading to $S_b=\sqrt{2}-1=0.414>0$. 

After Bell's original inequality (\ref{eq:bell}), Clauser {\it et al.} \cite{Clauser1969} extended Bell's analysis to cover realizable systems for experimental tests of quantum theory on nonlocality, leading to another famous form of Bell's inequality (namely the CHSH form of Bell's inequality, or the CHSH inequality),
\begin{equation}\label{eq:chsh}
S_c=|C(\alpha,\beta)+C(\alpha',\beta)|+|C(\alpha,\beta')-C(\alpha',\beta')|-2\le 0\ , 
\end{equation}
in which $\alpha'$ and $\beta'$ have the same meaning as $\alpha$ and $\beta$. The inequality (\ref{eq:chsh}) is violated by the correlation (\ref{eq:coscorr}) predicted by quantum theory with $\alpha=0$, $\alpha'=\pi/2$, $\beta=\pi/4$, and $\beta'=3\pi/4$ (in which case $S_c=2\sqrt{2}-2=0.828>0$), as verified by experiment \cite{weihs1998,Hensen2015}.

Later, Clauser and Horne derived a third form of Bell's inequality (the CH form of Bell's inequality, or the CH inequality), to address the practical issue of imperfect particle detection efficiency, \cite{Clauser1974,Eberhard1993},
\begin{eqnarray}\label{eq:ch}
 S_h&=&p_{+-}(\alpha,\beta)+p_{+-}(\alpha,\beta')+p_{+-}(\alpha',\beta)\nonumber\\
 &-&p_{+-}(\alpha',\beta')- p_+(\alpha)-p_-(\beta)\le 0\ ,
\end{eqnarray}
in which $p_{+}(\alpha)$ or $p_{-}(\beta)$ denotes the probability per measurement time $\tau$ of detecting a single electron $e_a$ or $e_b$ with a local parameter $\alpha$ or $\beta$, $p_{+-}(\alpha,\beta)$ is the joint probability of detecting the electron $e_a$ with a local parameter $\alpha$, and $e_b$ with $\beta$. The subscripts $+$ and $-$, respectively, indicate an outcome $+1$ for the detection of the electron $e_a$ and $-1$ for $e_b$. The actual values of $p_{+}(\cdot)$, $p_{-}(\cdot)$ and $p_{+-}(\cdot)$ depend on the measurement time $\tau$ and other practical parameters, but $p_{+-}(\alpha,\beta)=\varepsilon\ p'_{ab,AB}(\alpha,\beta)|_{A=-B}$, $p_{+}(\alpha)=\varepsilon p'_{+}(\alpha)$, and $p_{-}(\beta)=\varepsilon p'_{-}(\beta)$ are expected in practice. Here, $\varepsilon$ is a positive constant determined by the specific experimental parameters and $p'(\cdot)$ denotes the probability of the corresponding event per observation (measurement). Since $p'_{+}(\alpha)=p'_{-}(\beta)=1/2$, the CH inequality (\ref{eq:ch}) is violated by the joint probabilities of Eqs. (\ref{eq:jointq}) predicted by quantum theory (for example, with the parameters of $\alpha=0$, $\beta=\pi/4$, $\alpha'=\pi/2$ and $\beta'=-\pi/4$, one will have $S_h=2^{-1}(\sqrt{2}-1)\varepsilon=0.207\varepsilon>0$), which was verified in experiments \cite{Giustina2013,Christensen2013} with imperfect particle detection efficiencies of sufficiently high values (that is, detection efficiencies $\ge 2/3$) \cite{Eberhard1993}. 

Another version of inequality (the CH-E inequality, or the CH-E form of Bell's inequality) \cite{Eberhard1993,Bierhorst2015} was derived to solve the problem of memory effects (that is, the possibility of a local state varying over time with a possible dependence on earlier trials),
\begin{eqnarray}\label{eq:chB}
 S_e&=&p_{+-}(\alpha,\beta)-p_{+0}(\alpha,\beta')-p_{0-}(\alpha',\beta)\nonumber\\
 &&-p_{+-}(\alpha',\beta')\le 0\ ,\ \ \ 
\end{eqnarray}
where $p_{+-}(\cdot)$ stands for the joint probability per measurement time that both electrons are recorded with an outcome $+1$ for $e_a$ and $-1$ for $e_b$; $p_{+0}(\cdot)$ is the probability that electron $e_a$ is recorded with an outcome $+1$ but electron $e_b$ is not recorded, and $p_{0-}(\cdot)$ is the probability that electron $e_b$ is recorded with an outcome $-1$ but electron $e_a$ is not recorded. Noting that $p_{+0}(\alpha,\beta')=p_{0-}(\alpha',\beta)=0$ in quantum theory, one can trivially show that inequality (\ref{eq:chB}) is violated by the joint probabilities of Eqs. (\ref{eq:jointq}) (for example, the settings of $\alpha=0$, $\beta=\pi/4$, $\alpha'=\pi/2$ and $\beta'=-\pi/4$ lead to $S_e=4^{-1}\sqrt{2}\varepsilon=0.354\varepsilon>0$). Experiments \cite{Giustina2015,Shalm2015} that demonstrated this violation were reported a decade ago, claiming results that are free of loopholes.

\subsection{A local explanation of Bell violation based on the theory of vectorial probability}

All the above forms of Bell's inequality were derived starting from the same assumption of locality, from which physicists typically point out violation of Bell's theorem as nonlocality to explain why the experimental data do not match these inequalities \cite{salart2008,Giustina2013,Brunner2014,Giustina2015,Shalm2015,Hensen2015,bigbell2018}. In separated two-body systems, the experimental results of Bell violation were believed to rule out any local explanation of quantum-mechanical correlations, and most scientists were forced to admit either superluminal influences or extended physical entities that do not possess well-defined local properties \cite{Svetlichny1987}. Surely, the validity of claiming a nonlocal structure of quantum mechanics excludes all possibilities of explaining the results of the Bell test by a local theory. However, in the following, we show that Bell violation can be well described by the local theory developed based on the concept of vectorial probability and, therefore, a loophole of a new type exists in the Bell test.

\subsubsection{A local model in a four-dimensional probability space}

Following the theory described in the preceding context, a local model can be constructed and associated with a four-dimensional probability space $S$. In the model, we define a vectorial probability (per measurement time $\tau$) ruling over a stochastic particle detection event $z$ with binary results $Z=\pm 1$,
\begin{eqnarray} \label{eq:vpdef}
\mathbf{p}_{z,Z}(\xi)&=& 2^{-1}p_{z,Z}\Big[1+Z\cos\xi, 1-Z\cos\xi, \nonumber\\
&& \hspace{0.6in}Z\sin\xi, Z\sin\xi\Big] \ ,
\end{eqnarray} 
where $p_{z,Z}=\lVert\mathbf{p}_{z,Z}(\xi)\rVert$ is the traditional probability of the event observed with a result $Z$ in time $\tau$. The vectorial probability defined above can satisfactorily describe particle detection events in spin (or polarization) measurements in typical Bell experiments \cite{salart2008,Giustina2013,Brunner2014,Giustina2015,Shalm2015,Hensen2015,bigbell2018} and other experiments involving $m$-particle entanglement ($m\ge 3$) \cite{pan2000,Hamel2014}. We will first demonstrate how the local model based on the concept of vectorial probability can explain the results of the Bell test involving two entangled particles. To that end, setting $z=a$, $Z=-A$, and $\xi=\alpha$ for the stochastic events occurring in Alice's measurement, and $z=b$, $Z=B$, and $\xi=\beta$ for those in Bob's measurement. Then from Eq. (\ref{eq:vpdef}), we can obtain two vectorial probabilities,
\begin{eqnarray} \label{eq:vpchoice}
\mathbf{p}_{a,A}(\alpha)&=&2^{-1}p_{a,A}\Big[1-A\cos\alpha, 1+A\cos\alpha, \nonumber\\
&& \hspace{0.6in}-A\sin\alpha, -A\sin\alpha\Big]\ , \nonumber\\
\mathbf{p}_{b,B}(\beta)&=&2^{-1}p_{b,B}\Big[1+B\cos\beta, 1-B\cos\beta, \nonumber\\
&&\hspace{0.6in}B\sin\beta, B\sin\beta\Big] \ ,
\end{eqnarray} 
the length of each describing the likelihood of the corresponding particle detection event in the spin (polarization) measurement by Alice or Bob.

The orthogonality of the vectorial probabilities $\mathbf{p}_{a,A}(\alpha)$ and $\mathbf{p}_{b,B}(\beta)$ can be easily proved, that is, 
\begin{eqnarray} \label{eq:orthovp}
\mathbf{p}_{a,A}(\alpha)\cdot\mathbf{p}_{a,-A}(\alpha)=\mathbf{p}_{b,B}(\beta)\cdot\mathbf{p}_{b,-B}(\beta)&=&0 \ ,
\end{eqnarray} 
from which it follows that the probability of detecting a single electron $e_a$ (or $e_b$) is $p_a=p_{a,+1}+p_{a,-1}$ (or $p_b=p_{b,+1}+p_{b,-1}$). With Eqs. (\ref{eq:vpchoice}), it is trivial to show $p_{a,+1}=p_{a,-1}$ for the detection of the electron $e_a$ and $p_{b,+1}=p_{b,-1}$ for $e_b$. Consequently, $p_a=2p_{a,+1}=2p_{a,-1}$ and $p_b=2p_{b,+1}=2p_{b,-1}$, from which it follows that the probability (per measurement) of detecting the electron $e_a$ with each outcome $A$ is $p'_{a,A}=p_{a,A}/p_a=2^{-1}$ and the same is true for electron $e_b$, $p'_{b,B}=p_{b,B}/p_b=2^{-1}$. Then we can obtain the vectorial probabilities for detection of electrons per measurement (observation),
\begin{eqnarray} \label{eq:vpchoicep}
\mathbf{p}'_{a,A}(\alpha)&=&4^{-1}\Big[1-A\cos\alpha, 1+A\cos\alpha, \nonumber\\
&& \hspace{0.3in}-A\sin\alpha, -A\sin\alpha\Big]\ , \nonumber\\
\mathbf{p}'_{b,B}(\beta)&=&4^{-1}\Big[1+B\cos\beta, 1-B\cos\beta, \nonumber\\
&&\hspace{0.3in}B\sin\beta, B\sin\beta\Big] \ .
\end{eqnarray} 
Invoking Eqs. (\ref{eq:defcpbi}) and (\ref{eq:jointqco}), the joint probability of detecting both electrons with an outcome $A$ for the electron $e_a$ and $B$ for $e_b$ can be calculated as,
\begin{eqnarray} \label{eq:jointabe}
p'_{ab,AB}(\alpha,\beta)&=&\lVert\mathbf{p}'_{a,A}(\alpha)\rVert^{-1}\mathbf{p}'_{a,A}(\alpha)\cdot\mathbf{p}'_{b,B}(\beta) \nonumber\\
&=&4^{-1}\Big[1-AB\cos(\alpha-\beta)\Big]\ ,
\end{eqnarray} 
which exactly reproduces the results of Eqs. (\ref{eq:jointq}) and (\ref{eq:coscorr}) as predicted by quantum theory. 

\subsubsection{Entangled particles in a singlet state}

To show the connection between our local model and Bell's theorem, we consider the case in which the particle system is in a singlet state (a special Bell state). In this case, the local model turns out to be able to describe the statistical nature of the stochastic events in spin (or polarization) measurements. As such, the joint probability for the stochastic events ($a_1$ and $a_2$) produced in the detection of the two particles, $e_1$ and $e_2$, in the experiment is given by Eq. (\ref{eq:jointabe}), which leads to the correlation $C(\alpha,\beta)=-\cos(\alpha-\beta)$ according to Eq. (\ref{eq:coscorr}) and reproduce exactly what quantum theory predicts for the Bell experiment \cite{Bell1964}. As such, the local model predicts violation of any form of Bell's inequality in the same way as quantum mechanics does as long as two entangled particles are concerned.

\begin{figure} 
	\centering
	\includegraphics[width=0.45\textwidth]{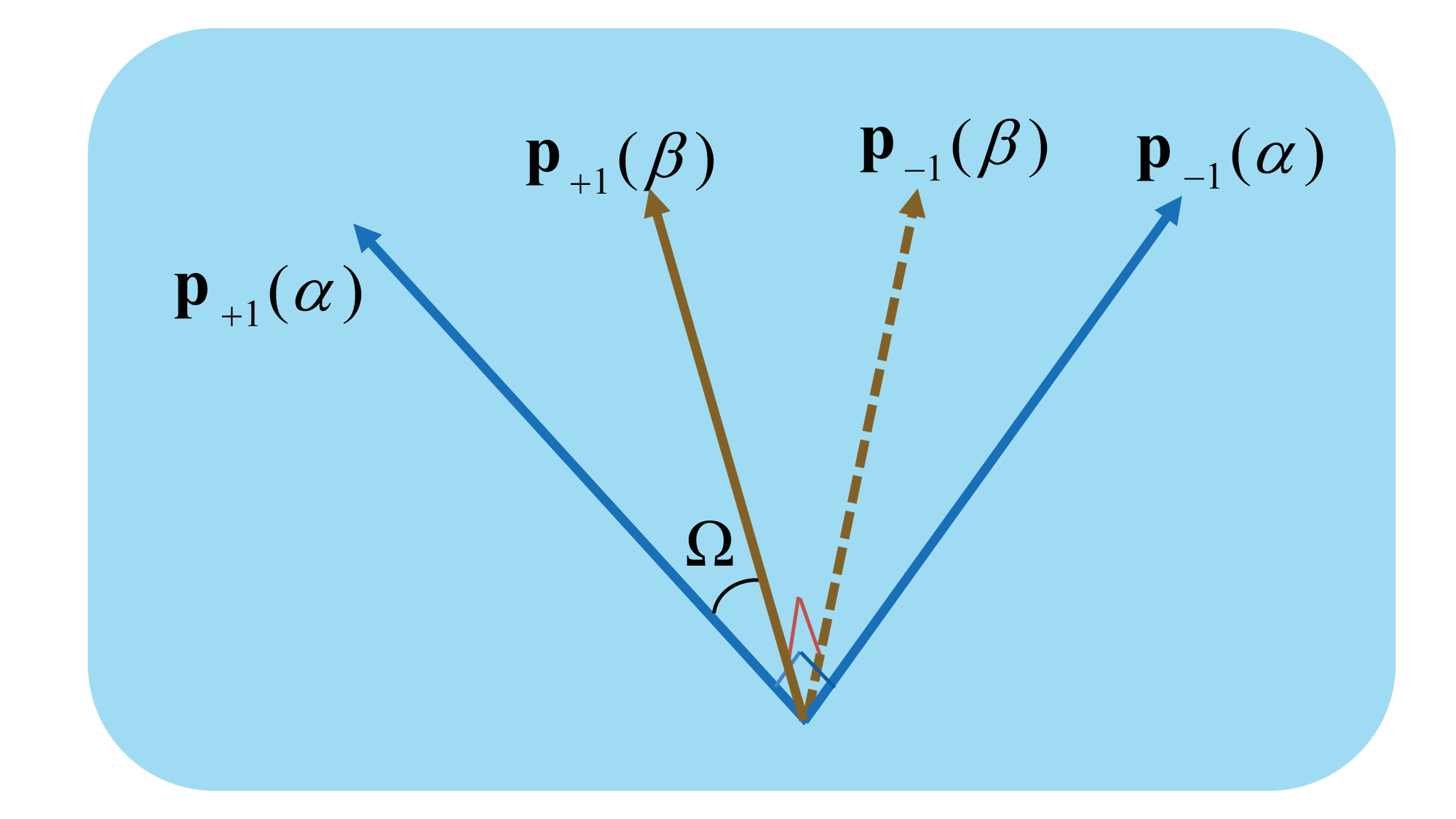} 

	\caption{Illustration of how the hidden geometry of vectorial probabilities works as the mechanism underlying the Bell violation. In a Bell test, the stochastic events of single-particle detection are described by one of the local vectors, $\mathbf{p}_{A}(\alpha)$ and $\mathbf{p}_{B}(\beta)$ ( $A=\pm1$ and $B=\pm1$). These four vectors are usually not coplanar in the four-dimensional probability space; the length of each vector is constant, but their directions are steered locally and independently by the variables $\alpha$ and $\beta$ respectively. This results in a variation of the angle $\Omega$ between $\mathbf{p}_{A}(\alpha)$ and $\mathbf{p}_{B}(\beta)$ and hence a change of the joint probability $p_{AB}(\alpha,\beta)$ as a function of $\alpha$ and $\beta$ given by Eq. (\ref{eq:coscorr}), giving rise to Bell violation.}
	\label{fig:jointdefp} 
\end{figure}

Despite its violation of Bell's theorem, our model has a local structure in the sense that the stochastic events in the spin (or polarization) measurement of each particle are ruled locally and independently by one of the vectorial probabilities, $\mathbf{p}_{A}(\alpha)$ or $\mathbf{p}_{B}(\beta)$, as shown by Eqs. (\ref{eq:vpchoice}); that is, the value of $\alpha$ does not influence the result of the event $b$, nor does the value $\beta$ affect the result of the event $a$. On the one hand, the event of detecting the electron $e_a$ is locally ruled by the parameter $\alpha$ through $\mathbf{p}_{A}(\alpha)$ and that of detecting the electron $e_b$ is governed by the parameter $\beta$ through $\mathbf{p}_{B}(\beta)$, that is, the two events are mutually independent no matter if they occur at separated places outside the cone of light. On the other hand, when the direction of $\mathbf{p}_{A}(\alpha)$ or $\mathbf{p}_{B}(\beta)$ is independently steered by the local parameter $\alpha$ or $\beta$, the angle $\Omega$ between $\mathbf{p}_{A}(\alpha)$ and $\mathbf{p}_{B}(\beta)$ varies, causing a variation in the joint probability $p_{AB}(\alpha,\beta)$ of the corresponding particle detection events according to Eq. (\ref{eq:defp}). It turns out that the joint probability and therefore the correlation of the events in the joint spin (or polarization) measurement changes as a function of $\alpha-\beta$ according to Eq. (\ref{eq:jointabe}) resulting from the otherwise hidden geometry of the vectorial probabilities, even though the local probability $p_{A}$ (or $p_{B}$) (per measurement) of detecting the electron $e_a$ (or $e_b$) with the result $A$ (or $B$) remains constant. This provides a descriptive explanation for how Bell violation comes into being in our local model, in which we do not need to introduce the concept of nonlocality.

\subsection{Bell violation with three entangled particles}

With a quantum system comprising three particles in a Greenberger-Horne-Zeilinger (GHZ) state, Bell's theorem can be violated by the correlation between the particle detection events produced in the joint measurement of the spin (or polarization) of the particles, with an inequality \cite{Svetlichny1987,Hamel2014} or without inequalities \cite{Greenberger1990,Bouwmeester1999,pan2000}. In this case, we will show that our local model can also reproduce the corresponding predictions of quantum theory and violate Bell's theorem in the same way as quantum mechanics.

To that aim, we generalize the local model leading to the result of Eq. (\ref{eq:jointabe}) to the situation of more entangled particles, in which a stochastic event can be produced by detecting one of the particles $e_1$, ..., and $e_m$ ($m\ge3$ is an integer), in a state with a wave function $|\psi_m\rangle$ of the form
\begin{eqnarray} \label{eq:mentan}
|\psi_m\rangle&=&2^{-1/2}\Big[|z_{s_1}\rangle_{e_1}\otimes ...|z_{s_j}\rangle_{e_j}...\otimes |z_{s_m}\rangle_{e_m}\nonumber\\ 
&&+s|z_{-s_1}\rangle_{e_1}\otimes ...|z_{-s_j}\rangle_{e_j}...\otimes |z_{-s_m}\rangle_{e_m}\Big] \ ,
\end{eqnarray} 
where $\otimes$ represents the tensor-product operation for vectors in a Hilbert space, $|z_{\pm s_j}\rangle_{e_j}$ ($j=1,...,m$) is the wave function of the particle $e_j$ with spin (or polarization) $\pm s_j$, the integer $s_j$ or $s$ has a binary value of $\pm 1$, and $s_j=\pm1$ respectively correspond to a particle $e_j$ with spin (or polarization) up and down along the $z$-axis in a 3D coordinate system. 

In the model, the measurement of the spin (or polarization) of the particle $e_j$ is indicated by $\sigma_j$, with a probabilistic measurement result of $A_{j}=\pm 1$. Each measurement $\sigma_j$ is characterized by a polar angle $\alpha_j$ and an azimuthal angle $\xi_j$ that define the direction along which the spin (polarization) of the particle is measured. Then each particle $e_j$ of the entangled system in the state given by Eq. (\ref{eq:mentan}) can be detected with a result $A_j$, producing a stochastic event $a_j$ ruled by one of the corresponding vectorial probabilities \cite{feng20252}, 
\begin{eqnarray} \label{eq:vpchoicem}
\mathbf{p}_{A_1}(\sigma_1)&=&4^{-1}\Big[(1+s_1A_1\cos\alpha_1), (1-s_1A_1\cos\alpha_1),  \nonumber\\
&&\hspace{0.3in}sA_1\sin\alpha_1e^{i\xi_1}, sA_1\sin\alpha_1e^{-i\xi_1}\Big]\ ,\ \mbox{or}\nonumber\\
\mathbf{p}_{A_l}(\sigma_l)&=&4^{-1}\Big[(1+s_lA_l\cos\alpha_l), (1-s_lA_l\cos\alpha_l),  \nonumber\\
&&\hspace{0.3in}  A_l\sin\alpha_le^{i\xi_l}, A_l\sin\alpha_le^{-i\xi_l}\Big] \ ,
\end{eqnarray} 
for $l=2,...,\ \mbox{or}\ m$. It is trivial to show that the probability of detecting the particle $e_j$ with a result of $A_j$ is equal to $p_{A_j}(\sigma_j)=\lVert\mathbf{p}_{A_j}(\sigma_j)\rVert=1/2$ ($j=1,...,m$) and that $\mathbf{p}_{A_j}(\sigma_j)\cdot \mathbf{p}^*_{-A_j}(\sigma_j)=0$, in which $\mathbf{p}^*(\cdot)$ is the conjugate complex of $\mathbf{p}(\cdot)$. 

The above results show that, as long as the stochastic event generated in the detection of a single particle is concerned, the local model agrees with quantum theory. In the following, we will prove that the model also reproduces the prediction of quantum mechanics when more particles are simultaneously measured. According to the model and the local theory developed above, the joint probability that a spin (or polarization) measurement on the entangled system $|\psi_m\rangle$ generates an outcome of $A_1$, ..., $A_j$, ..., and $A_m$ reads
\begin{eqnarray} \label{eq:mentanp}
 &&p_{A_1...A_m}(\sigma_1,...,\sigma_m) \nonumber\\
 &=&\lVert\mathbf{p}_{A_1}(\sigma_1)\rVert^{-m+1}\Big[\mathbf{p}_{A_1}(\sigma_1)\cdot\ ...\nonumber\\
&&\hspace{0.8in} \cdot\mathbf{p}_{A_j}(\sigma_j)\cdot...\cdot\mathbf{p}_{A_m}(\sigma_m)\Big]\nonumber\\
&=& \lVert\mathbf{p}_{A_1}(\sigma_1)\rVert^{-m+1} \sum_{k=1}^4p_{A_1;k}(\sigma_1)\times...\nonumber\\
&&\hspace{0.6in}\times p_{A_j;k}(\sigma_j)\times...\times p_{A_m;k}(\sigma_m)
\ ,
\end{eqnarray}
in which the mathematical operation of the successive-dot product (s-dot product) was invoked for the vectors in the calculation, and $p_{A_j;k}(\sigma_j)$ are the projected components of $\mathbf{p}_{A_j}(\sigma_j)$, respectively, along the unity base vectors $\hat{s}_k$ in a four-dimensional probability space $S_4$. The values of these components are given by Eqs. (\ref{eq:vpchoicem}); for example, $p_{A_1;2}(\sigma_1)=4^{-1}(1-s_1A_1\cos\alpha_1)$ and $p_{A_1;3}(\sigma_1)=4^{-1}sA_1\sin\alpha_1e^{i\xi_1}$.

\subsubsection{Violation of a local inequality for three-particle systems}

In the following, we show that a local inequality for three-particle systems in Bell's theorem can be violated by our local model in the same way as quantum mechanics does. The inequality (namely the Svetlichny inequality) \cite{Svetlichny1987} was derived as a generalization of Bell's original inequality \cite{Bell1964} to cover the situation in which stochastic particle detection events are produced in dichotomous spin (polarization) measurements on each part of a three-particle system (for example, electron $e_1$, $e_2$ and $e_3$). The derivation of this inequality was based on a statistical model as described by the following correlation function,
\begin{equation} \label{eq:lmodel3}
    C(\sigma_1,\sigma_2,\sigma_3)=\int \mbox{d}\chi\ \rho(\chi) u(\chi;\sigma_1,\sigma_2)v(\chi;\sigma_3),
\end{equation}
where $\sigma_1$, $\sigma_2$ and $\sigma_3$, respectively, characterize the measurements on the corresponding particles, and $|u(\chi;\sigma_1,\sigma_2)|\le 1$ and $|v(\chi;\sigma_3)| \le 1$, both of which can be understood as probability densities conditioned to the variable $\chi$. In the model, the result of a dichotomous measurement on the subsystem comprising $e_1$ and $e_2$ is given by the sign of $u(\chi;\sigma_1,\sigma_2)$ (that is, $+1$ if $u(\chi;\sigma_1,\sigma_2)> 0$ and $-1$ if $u(\chi;\sigma_1,\sigma_2)< 0$), and the result of a measurement on $e_3$ is determined by the sign of $v(\chi;\sigma_3)$. 

The derived Svetlichny inequality is \cite{Svetlichny1987,Hamel2014}
\begin{eqnarray}\label{eq:lmodel3}
S_{\rm svet}&\equiv&|C(\sigma_1,\sigma_2,\sigma_3)+C(\sigma_1,\sigma_2,\sigma'_3)+C(\sigma_1,\sigma'_2,\sigma_3)\nonumber\\
&&+C(\sigma'_1,\sigma_2,\sigma_3)-C(\sigma_1,\sigma'_2,\sigma'_3)-C(\sigma'_1,\sigma_2,\sigma'_3)\nonumber\\
&&-C(\sigma'_1,\sigma'_2,\sigma_3)-C(\sigma'_1,\sigma'_2,\sigma'_3)|-4\le 0\ .
\end{eqnarray}
It can be shown \cite{Hamel2014} that the Svetlichny inequality can be violated by an experiment involving spin (or polarization) measurements on a three-electron (or three-photon) system in the GHZ state as given by Eq. (\ref{eq:mentan}) for $m=3$ \cite{Greenberger1990,Bouwmeester1999,pan2000}. To test the inequality, the spin measurement settings were chosen such that $\sigma_1=\sigma_2=\hat{s}_x$, $\sigma'_1=\sigma'_2=\hat{s}_y$, $\sigma_3=2^{-1/2}(\hat{s}_x-\hat{s}_y)$ and $\sigma'_3=2^{-1/2}(\hat{s}_x+\hat{s}_y)$ \cite{Hamel2014}, where $\hat{s}_x$ and $\hat{s}_y$, respectively, represent spin measurements along the $x$ and $y$ directions. With these settings, quantum theory predicts $S_{\rm svet}=4(\sqrt{2}-1)>0$ which violates the Svetlichny inequality, as confirmed in experimental observations using a three-photon system in a GHZ state \cite{Hamel2014}.

Although the violation of the Svetlichny inequality was claimed to guaranty the presence of multipartite nonlocality \cite{Svetlichny1987,Hamel2014}, we will prove in the following that the violation of the inequality can also be well explained with the local model of Eqs. (\ref{eq:vpchoicem}) and (\ref{eq:mentanp}). For a system in a GHZ state, we have $s_1=s_2=s_3=+1$ and $s=1$ for the two equations. In accordance with this model, the joint probability of particle detection events in spin (polarization) measurements ($\sigma_1$, $\sigma_2$, and $\sigma_3$) is
\begin{eqnarray} \label{eq:jointabcspin}
&&p'_{e_1e_2e_3,A_1A_2A_3}(\sigma_1,\sigma_2,\sigma_3)\nonumber\\
 &=&\lVert\mathbf{p}'_{e_1,A_1}(\sigma_1)\rVert^{-2} \Big[\mathbf{p}'_{e_1,A_1}(\sigma_1)\cdot \mathbf{p'}_{e_2,A_2}(\sigma_2)\cdot \mathbf{p}'_{e_3,A_3}(\sigma_3)\Big]\nonumber\\
&=& 16^{-1}\Big[(1+A_1\cos\alpha_1)(1+A_2\cos\alpha_2)(1+A_3\cos\alpha_3)\nonumber\\
&&\ \ +(1-A_1\cos\alpha_1)(1-A_2\cos\alpha_2)(1-A_3\cos\alpha_3)\nonumber\\
&&\ \ +2s\ A_1A_2A_3\sin\alpha_1\sin\alpha_2\sin\alpha_3\cos(\zeta_1+\zeta_2+\zeta_3)\Big] \ , \nonumber\\
\end{eqnarray}
with which it is trivial to confirm that
\begin{eqnarray} \label{eq:corrabcspin}
&&C(\sigma_1,\sigma_2,\sigma_3)\nonumber\\
&=&\sum_{A_1A_2A_3=-1}^{+1} A_1A_2A_3\ 
p'_{e_1e_2e_3,A_1A_2A_3}(\sigma_1,\sigma_2,\sigma_3)\nonumber\\
&=&\sum_{A_1A_2A_3=-1}^{+1} \frac{A_1A_2A_3}{8}\big[1+sA_1A_2A_3\cos(\zeta_1+\zeta_2+\zeta_3)\big] 
\nonumber\\
&=& s\ \cos(\zeta_1+\zeta_2+\zeta_3)\big]
\nonumber\\
 &=&2^{-1/2}s \ .
\end{eqnarray}
Similarly, the values of the last four correlation functions in Eq. (\ref{eq:lmodel3}) can be calculated as $-2^{-1/2}s$ and each of the other terms as $2^{-1/2}s$. In the calculations, note that $\alpha_{1,2,3}=\pi/2$; and $\zeta=0$ for a measurement $\hat{s}_x$, $\zeta=\pi/2$ for a measurement $\hat{s}_y$, $\zeta=-\pi/4$ for a measurement $2^{-1/2}(\hat{s}_x-\hat{s}_y)$, and $\zeta=\pi/4$ for a measurement $2^{-1/2}(\hat{s}_x+\hat{s}_y)$. As such, since $|s|=1$, the result of $S_{\rm svet}=4(\sqrt{2}-1)>0$ is obtained, which exactly reproduces the prediction of quantum theory, for a violation of the Svetlichny inequality.

\subsubsection{Violation of Bell's theorem without inequalities}

The entanglement of $m$ particles ($m\ge 3$) was believed to cause a conflict with all local models for non-statistical predictions of quantum theory \cite{Greenberger1990,Bouwmeester1999}, without the need of any inequality derived from a statistical model. This was the original incentive to make use of three-particle systems in a GHZ state for the violation of Bell's theorem as evidence of nonlocality. To illustrate the idea, consider a particular three-particle state of the form \cite{pan2000},
\begin{eqnarray} \label{eq:mentan3}
|\psi_3\rangle&=&2^{-1/2}\Big[|{+1_1}\rangle_{z}\otimes |{+1_2}\rangle_{z}\otimes |{+1_3}\rangle_{z}\nonumber\\ 
&&\hspace{0.3in}+|{-1_1}\rangle_{z}\otimes|{-1_2}\rangle_{z}\otimes |{-1_3}\rangle_{z}\Big] \ ,
\end{eqnarray}
wherein $|{\pm 1}_j\rangle_{z}$ ($j=1,2,3$) are the eigenstates of the operator $\hat{s}_z$ describing the spin (polarization) measurement on the particle $e_j$ along the $z$-direction. 

Mathematically, the wave function $|\psi_3\rangle$ of the three-particle state given by Eq. (\ref{eq:mentan3}) can also be expressed in terms of the eigenstates of other operators $\hat{s}_x$ and $\hat{s}_y$, respectively describing the spin (polarization) measurement on the particle along the $x$-direction and the $y$-direction. These expressions for the wave function $|\psi_3\rangle$ show that, in each of the joint spin measurements $\hat{s}_y\hat{s}_y\hat{s}_x$, $\hat{s}_y\hat{s}_x\hat{s}_y$ and $\hat{s}_x\hat{s}_y\hat{s}_y$, respectively, on the three particles, all possible results of the individual measurement on each particle can be predicted with certainty based on the results of the measurements on the other two \cite{pan2000}. In this measurement scenario, the local argument-based Einstein locality is understood as follows: no information-carrying signal can travel faster than the speed of light, which demands that any particular spin measurement result achieved for any particle should not depend on the specific measurements that are performed on the other two particles or on their outcomes. This understanding can be directly tested in experiments with a three-particle system in a GHZ state as follows.

For a spin (polarization) measurement $\hat{s}_x$ or $\hat{s}_y$ on a particle $j$, the result can be denoted as $X_j$ or $Y_j$ ($j=1,2,3$). With the wave function of the three-particle GHZ state given by Eq. (\ref{eq:mentan3}), it can be shown from quantum theory that joint measurements on the system will lead to the following outcomes: $Y_1Y_2X_3=-1$ for a measurement $\hat{s}_y\hat{s}_y\hat{s}_x$, $Y_1X_2Y_3=-1$ for a measurement $\hat{s}_y\hat{s}_x\hat{s}_y$, and $X_1Y_2Y_3=-1$ for a measurement $\hat{s}_x\hat{s}_y\hat{s}_y$ \cite{pan2000}. Now, if another joint spin (polarization) measurement $\hat{s}_x\hat{s}_x\hat{s}_x$ is carried out on the system, the corresponding result should be $X_1X_2X_3=(X_1Y_2Y_3)(Y_1X_2Y_3)(Y_1Y_2X_3)$ due to Einstein locality \cite{pan2000}, according to which any specific measurement $\hat{s}_x$ or $\hat{s}_y$ on one particle must be independent of the measurements on other particles. Since $Y_j=\pm 1$ ($j=1,2,3$), one has $Y_1^2=Y_2^2=Y_3^2=1$, resulting in $X_1X_2X_3=-1$. The only possibility to achieve this result is that the state $|\psi_3\rangle$ of the system after measurement $\hat{s}_x\hat{s}_x\hat{s}_x$ collapses into $|x_{-1}\rangle_{e_1}\otimes |x_{-1}\rangle_{e_2}\otimes |x_{-1}\rangle_{e_3}$, $|x_{-1}\rangle_{e_1}\otimes |x_{+1}\rangle_{e_2}\otimes |x_{+1}\rangle_{e_3}$, $|x_{+1}\rangle_{e_1}\otimes |x_{-1}\rangle_{e_2}\otimes |x_{+1}\rangle_{e_3}$ or $|x_{+1}\rangle_{e_1}\otimes |x_{+1}\rangle_{e_2}\otimes |x_{-1}\rangle_{e_3}$. However, from the wave function of Eq. (\ref{eq:mentan3}), it follows that none of the above four potential possibilities is allowed by quantum theory; instead, the system can only be projected by measurement in the state of $|x_{+1}\rangle_{e_1}\otimes |x_{+1}\rangle_{e_2}\otimes |x_{+1}\rangle_{e_3}$, $|x_{+1}\rangle_{e_1}\otimes |x_{-1}\rangle_{e_2}\otimes |x_{-1}\rangle_{e_3}$, $|x_{-1}\rangle_{e_1}\otimes |x_{+1}\rangle_{e_2}\otimes |x_{-1}\rangle_{e_3}$ or $|x_{-1}\rangle_{e_1}\otimes |x_{-1}\rangle_{e_2}\otimes |x_{+1}\rangle_{e_3}$, as confirmed by experiment \cite{pan2000}.

From the experimental verification of the prediction of quantum theory, a contradiction between locality and quantum theory was claimed with the conclusion that whenever a local model predicts a particular result of a measurement on one electron based on the results for the others, quantum theory certainly predicts the opposite result \cite{pan2000}. However, in the following, we will show that our local model, specifically Eq. (\ref{eq:jointabcspin}), predicts the same results as quantum theory, as follows. 

For the spin (polarization) measurement $\hat{s}_y\hat{s}_y\hat{s}_x$ (that is, with the settings of $\sigma_1=\hat{s}_y$, $\sigma_2=\hat{s}_y$ and $\sigma_3=\hat{s}_x$), one has $\alpha_{1,2,3}=\pi/2$, $\zeta_{1,2}=\pi/2$, and $\zeta_3=0$. Plugging these values and $s=1$ into Eq. (\ref{eq:jointabcspin}) leads to 
\begin{equation} \label{eq:yyx}
    p'_{e_1 e_2 e_3,Y_1 Y_2 X_3}=8^{-1}(1-Y_1 Y_2 X_3)\ .
\end{equation}
The result of Eq. (\ref{eq:yyx}) shows that the joint probability of particle detection in the measurement $\hat{s}_y\hat{s}_y\hat{s}_x$ is $4^{-1}$ for the result $Y_1 Y_2 X_3=-1$ and $0$ for $Y_1 Y_2 X_3=+1$. Similarly, for the measurements $\hat{s}_y\hat{s}_x\hat{s}_y$ and $\hat{s}_x\hat{s}_y\hat{s}_y$, respectively, one can have
\begin{eqnarray} \label{eq:yxy}
p'_{e_1 e_2 e_3,Y_1 X_2 Y_3}&=&8^{-1}(1-Y_1 X_2 Y_3)\nonumber\\
p'_{e_1 e_2 e_3,X_1 Y_2 Y_3}&=&8^{-1}(1-X_1 Y_2 Y_3)\ ,
\end{eqnarray} 
which show that the joint probability of particle detection in the measurement $\hat{s}_y\hat{s}_x\hat{s}_y$ is $4^{-1}$ for $Y_1 X_2 Y_3=-1$ and $0$ for $Y_1 X_2 Y_3=+1$; the joint probability of particle detection in the measurement $\hat{s}_x\hat{s}_y\hat{s}_y$ is also $4^{-1}$ for $X_ 1Y_2 Y_3=-1$ and $0$ for $X_1 Y_2 Y_3=+1$. All the above predictions by our local model agree well with quantum theory and experiment. 

Now, we inspect what our local model predicts for the measurement $\hat{s}_x\hat{s}_x\hat{s}_x$, for which we have $\alpha_{1,2,3}=\pi/2$ and $\zeta_{1,2,3}=0$. From Eq. (\ref{eq:jointabcspin}), it follows that 
\begin{equation} \label{eq:xxx}
    p'_{e_1 e_2 e_3,X_1 X_2 X_3}=8^{-1}(1+X_1 X_2 X_3)\ ,
\end{equation}
which shows that the joint probability in the measurement $\hat{s}_x\hat{s}_x\hat{s}_x$ is $4^{-1}$ for the result $X_1 X_2 X_3=+1$ and $0$ for $X_1 X_2 X_3=-1$, which is exactly what quantum theory predicts, as verified by experimental observation \cite{pan2000}. In our local model, the individual spin (polarization) measurement on each particle depends only on the state of the local particle and is independent of the states of other particles. To be specific, the local event of particle detection in the spin (polarization) measurement on each particle $e_j$ ($j=1,...,m$) is governed by its own vectorial probability $\mathbf{p}'_{e_j,A_j}(\sigma_j)$ alone, which does not depend on any local parameters $\sigma_l$ or $s_l$ associated with other particles $e_l$ ($l\ne j$, $l=1,...,m$) according to Eqs. (\ref{eq:vpchoicem}). As such, the violation of Bell's theorem by the correlations of the stochastic events occurring in the joint spin (or polarization) measurement on the three-particle system in the GHZ state can be explained by our local model, which constitutes a loophole in the Bell test of relevance.

\section{Discussions}

Despite its locality, our model violates Bell's theorem no matter whether it is with an inequality (the CHSH inequality, the CH inequality, the CH-E inequality, or the Svetlichny inequality) or without inequalities, as proved above. The model allows for an explanation of the previous results in the Bell test by the local theory developed based on the concept of vectorial probability and, therefore, proves the presence of a new kind of loophole (the vectorial probability loophole) that remains open in the Bell test. However, closing this loophole is not a trivial task. Comprehensive investigations are highly needed in theory to find some statistical property of the stochastic events occurring in quantum measurements to distinguish the predictions of quantum mechanics and those of the local theory of vectorial probability. Alternatively speaking, we need to find out how to create stochastic events whose statistical property, for sure, can be fully described by quantum mechanics but, in contrast, must be beyond description in the theory of vectorial probability. We leave this task to be accomplished in the future.

In history, probability theory based on the concept of scalar probability was assumed to be a solid statistical foundation on which quantum theory has been formulated. A typical example is Born's statistical interpretation of the wave function, according to which the squared module of a normalized wave function describing a quantum object is equal to the (scalar) probability that the object appears in certain circumstance or a result comes up in a quantum measurement. This interpretation encounters no problem for a single quantum object or for a measurement on it. However, when it comes to the joint spin (or polarization) measurement of two or more particles in an entangled state, Born's statistical interpretation (together with probability theory) cannot provide a satisfactory explanation for the Bell violation, which consequently has forced physicists to accept it as evidence of the nonlocality of complex quantum systems \cite{Bell1964,Clauser1969,Freedman1972,Clauser1974,Fry1976,Aspect1982,Eberhard1993,weihs1998,Giustina2013,Bierhorst2015,Giustina2015,Shalm2015,Hensen2015}. 

After continuous efforts of theoretical and experimental study, nonlocality has been accepted as a fundamental element of quantum mechanics. However, doubts about nonlocality persist \cite{Matzkin2008,Khrennikov2016,Kupczynski2017,Pons2017,feng20253}. One of the key reasons why the debate has persisted for many decades is a dearth of a unified framework with a solid foundation in which all potential loopholes in the Bell test can be reliably recognized. The results presented in this work, together with the identification of the detection loophole beyond unfair sampling in the previous work \cite{tu2026}, has sounded a timely alarm about the lack of such a framework in theory.

Although we cannot close the vectorial probability loophole in the Bell test at this moment, the revelation of its existence is still of great interest for the fundamental study in quantum physics and other fields of relevance. First of all, the local model constructed on the concept of vectorial probability cannot in any way be fully covered by Born's statistical interpretation that was proposed on the concept of scalar probability. The point will become more obvious if one is aware that the quantum theory (which is based on Born's interpretation) can only predict Bell violation as a result of measurement but cannot explain the physical mechanism underlying this result without introducing an extra element (for example, the concept of nonlocality). It would be interesting to integrate the concept of vectorial probability into quantum mechanics for a generalized version of the statistical interpretation of the wave function. Thus, a one-to-one mapping can be established between the wave function of a quantum object undergoing a certain measurement and the vectorial probability of the stochastic events produced in the corresponding measurement. Secondly, quantum correlations that violate Bell's theorem are of great importance in the development of advanced quantum technologies, such as quantum information processing \cite{nielsen2000,Barrett2005,Gerhardt2011,Gerhardt2011p,Christensen2013,benenti2019,Zia2023,Liu2024,Cao2024,YLi2025}. The advantages of quantum technologies over their classical analogs, which were previously attributed to nonlocality, may be better understood and quantitatively described on the basis of the theory of vectorial probability. Last but not least, at this moment, no one can show how the vectorial probability loophole in the Bell test can eventually be closed. If quantum mechanics turns out to have a local structure, then the theory of vectorial probability can provide the necessary mathematical tool to restore locality to quantum theory. As such, a solid foundation can be established on the quantity of vectorial probability to formulate the basic concepts of entanglement \cite{Almeida2007,horodecki2009,Tendick2020} and quantum steering \cite{Schrodinger1935,Walborn2011,reid2019,uola2020}, and to better understand the physical procedure of quantum measurement \cite{Braginsky1992,Wiseman2009,Busch2016} and decoherence \cite{Zurek2003,Wilkinson2020}. 

As a new concept in mathematics, vectorial probability will also be of great essence for relevant research in the fields where many other important phenomena may be studied in depth beyond the scope of the conventional concept of scalar probability. For example, in the context of the quantum Markov chain \cite{Dhahri2023}, a quantum mechanical system that experiences transitions from one state to another according to certain probability rules, entanglement and quantum correlations have a profound impact on the evolution and behavior of the system. In this scenario, the concept of vectorial probability may provide a promising way to understand the overall dynamics and information flow within the system. Another example is Bayes' theorem, which is formulated with the concept of conditional probability. It turns out that vectorial probability can serve as an alternative mathematical basis on which the conditional probability formula is expressed. To illustrate, consider two stochastic events, $a$ and $b$, and the probability $p(a|b)$ of event $a$ occurring given event $b$. The conditional probability formula is mathematically expressed as
\begin{equation}
    p(a|b)=p(a\cap b)/p(b)\ ,
\end{equation}
in which $p(a\cap b)$ is the joint probability that both events occur. Using the notation of vectorial probability, the conditional probability formula can be re-expressed in the form of 
\begin{eqnarray} \label{eq:condip}
    p(a|b)&=&\mathbf{p}(a)\cdot \mathbf{p}(b)/p(b)\nonumber\\
    &=&\mathbf{p}(a)\cdot \mathbf{p}_0(b)\ ,
\end{eqnarray}
where $\mathbf{p}(a)$ or $\mathbf{p}(b)$ denotes the vectorial probability of the event $a$ or $b$, and $\mathbf{p}_0(b)$ is a unit vector along the direction of $\mathbf{p}(b)$ in the probability space. The formula (\ref{eq:condip}) suggests that the conditional probability $p(a|b)$ is equal to the projected component of $\mathbf{p}(a)$ along the direction of $\mathbf{p}(b)$ in the probability space; the geometry of the conditional probability $p(a|b)$, which was previously unknown, is revealed by the quantity of vectorial probability. 

In practical applications of Bayes' theorem, the conditional probability $p(a|b)$ or $p(b|a)$ may evolve over time depending on some uncontrollable parameters. Suppose that the two vectorial probabilities of events $a$ and $b$ are, respectively, independently dictated by two parameters $\alpha$ and $\beta$. Then, according to Eq. (\ref{eq:condip}), the corresponding conditional probability should also be ruled by these parameters, that is,
\begin{eqnarray} \label{eq:condipab}
    p(a|b;\alpha,\beta)&=&\mathbf{p}(a;\alpha)\cdot \mathbf{p}(b;\beta)/\lVert \mathbf{p}(b;\beta)\rVert\ \mbox{or}\nonumber\\
    p(b|a;\alpha,\beta)&=&\mathbf{p}(a;\alpha)\cdot \mathbf{p}(b;\beta)/\lVert \mathbf{p}(a;\alpha)\rVert\ .
\end{eqnarray}
From Eqs. (\ref{eq:condipab}), it follows that the vectorial probabilities carry more information about the two stochastic events than the conditional probabilities, in the sense that the evolution of the conditional probabilities can be uniquely determined if $\mathbf{p}(a;\alpha)$ and $\mathbf{p}(b;\beta)$ are known as functions of $\alpha$ and $\beta$.

\section{Conclusions}

We have recognized in theory a new loophole (vectorial probability loophole) in the Bell test, based on a local theory developed with the concept of vectorial probability, which is introduced into the probability theory as a result of inspiration by an experiment that demonstrates the possibility of violating the CH-E inequality by classical correlations of stochastic events. The new concept, quantified as a vector with interesting but hidden geometry, is a generalization of the conventional concept of scalar probability from the one-dimensional probability space to high dimensions. In the local theory developed here, we have shown how the probability of a single stochastic event and the joint probability of more than one event can be quantitatively described by the new mathematical concept. Then we have applied our local theory to describe the correlations of stochastic quantum events that occur in the Bell test and shown that previously known results of Bell violation can be well explained with our local theory based on the concept of vectorial probability. The local theory can reproduce exactly what quantum mechanics predicts about the violation of Bell's theorem when up to three entangled particles are concerned, which proves the existence of the vectorial probability loophole in the Bell test and raises a timely alarm about the dearth of a united framework with sound base to fully recognize all possible loopholes. Although we have not found a feasible way to close the loophole, the identification of it in the Bell test will be of great interest and of fundamental importance in both the study of quantum science and the development of advanced quantum technologies.

\section*{Acknowledgments}
This work was financially supported by the National Natural Science Foundation of China (grant number 12074110) and Hubei Polytechnic University University-Level Scientific Research Projects (grant number 25xjz01Y).

\section*{Notes}
The authors declare that they have no competing interests.

\section*{Data and materials availability}
The data required to support or replicate the claims made in the article are available at https://www.scidb.cn/s/Y3YRZr .


\bibliography{apssamp}


\end{document}